\documentclass{aa}  
\usepackage{graphicx}
\usepackage{txfonts}
\usepackage[dvipsnames]{xcolor}
\usepackage[normalem]{ulem}
\usepackage{array, multirow, boldline}
\usepackage{multirow}
\usepackage{parskip}
\usepackage{float}
\usepackage{caption}
\usepackage{placeins}
\usepackage{threeparttable} 
\usepackage{booktabs} 
\usepackage{needspace}

\usepackage[colorlinks,linkcolor=MidnightBlue,citecolor=MidnightBlue,linktocpage=true,breaklinks, 
plainpages=false,urlcolor=MidnightBlue]{hyperref}
\def\thethree{{\sc The Three Hundred Project}}

\usepackage[]{subfig}
\begin{document} 

   \title{Probing the baryonic--dark matter connection in galaxy clusters using X-rays with gated recurrent unit neural networks}


   \author{Asif Iqbal
          \inst{1}
          \and 
           Subhabrata Majumdar\inst{2}
          \and
           Weiguang Cui\inst{3}
          \and
          Elena Rasia\inst{4,5,6}
          \and 
          Gabriel W. Pratt \inst{7}  
          \and 
          Daniel de Andres\inst{8}    
          }
          
   \institute{Universit\'e de Lille, URePSSS - Unit\'e de Recherche Pluridisciplinaire Sport Sant\'e Soci\'et\'e, F-59000 Lille, France
              \email{asif-iqbal.ahangar@univ-lille.fr}
         \and
        Tata Institute of Fundamental Research, 1 Homi Bhabha Road, Colaba, Mumbai 400005, India
        \and 
         Departamento de F\'isica Te\'{o}rica, M\'{o}dulo 15 Universidad Aut\'{o}noma de Madrid, 28049 Madrid, Spain
           \and 
            INAF - Osservatorio Astronomico di Trieste, via Tiepolo 11, I-34131 Trieste, Italy
           \and IFPU, Via Beirut, 2, 3I-4151 Trieste, Italy
           \and
           Department of Physics; University of Michigan, Ann Arbor, MI 48109, USA
        \and 
        Universit\'e Paris-Saclay, Universit\'e Paris Cit\'e CEA, CNRS, AIM, 91191, Gif-sur-Yvette, France 
        \and
          Non-linear Dynamics, Chaos and Complex Systems Group, 
        Departamento de Geología, Física y Química Inorgánica, 
        Universidad Rey Juan Carlos, 
        Tulipán s/n, 28933 Móstoles, Madrid, Spain
             }
   
   \titlerunning{Galaxy Clusters mass profiles from X-rays using deep learning}

 
  \abstract
{Precise mass measurements are essential for cosmology, but traditional hydrostatic equilibrium (HSE) methods are prone to bias, particularly in dynamically disturbed systems. In this work, we introduce a deep learning approach based on a novel gated recurrent unit (GRU) that predicts the total three-dimensional mass profiles of galaxy clusters directly from the spherically symmetric radial profiles of the intracluster medium (ICM), including temperature, density, pressure, and enclosed gas mass. Treating these radial profiles as sequential data allows the GRU to capture physically motivated dependencies across radii while naturally handling profiles with different numbers of radial bins. We trained and validated our model on high-resolution hydrodynamical simulations from \thethree, achieving a $1\sigma$ scatter of $\sim5\%$ in cluster mass predictions from idealised three-dimensional profiles over most of the cluster region, which is substantially lower than what would be obtained with conventional HSE mass estimates. The model also provides robust radius-dependent uncertainty estimates, while observational uncertainties can be incorporated through a Monte Carlo approach. Our algorithm demonstrates robustness against variations in data quality and cluster morphology. We further demonstrate that the model generalises well across independent hydrodynamical simulation suites (GIZMO-SIMBA and GADGET-X) when trained jointly, recovering unbiased mass profiles in both cases. Feature importance analysis shows that enclosed gas mass dominates the predictions, with pressure and temperature shaping the radial structure, reflecting physically meaningful mass assembly. To examine mass estimates for observational data, we applied the GRU approach to the REXCESS and X-COP galaxy cluster samples observed with XMM-Newton. The GRU and HSE mass estimates show close agreement for the lower-mass REXCESS sample, while HSE masses are systematically lower than the GRU predictions for the higher-mass X-COP sample, an offset consistent with hydrostatic mass bias. This approach provides a powerful data-driven framework for precise and unbiased cluster mass estimation, bridging simulations and observations to improve the accuracy of cluster-based cosmological studies. Moreover, it can be readily extended to Sunyaev–Zel'dovich and optical datasets, making it a versatile tool for multi-wavelength cluster mass inference.
}

   \maketitle
%
\section{Introduction}
Galaxy clusters provide unparalleled insight into the formation and evolution of the cosmic web \citep{2004ApJ...613...41M,2011ARA&A..49..409A,2024A&A...690A.238A,2025A&A...703L..18A}. The vast majority of the mass in these colossal assemblies is composed of dark matter, a mysterious and invisible substance constituting approximately 80-90\% of the Universe's total matter content. Interwoven with this unseen component is the intracluster medium (ICM), a dynamic reservoir of hot ionised gas that emits X-rays and constitutes the majority of the baryonic content of the clusters \citep{2019SSRv..215...25P,2022A&A...662A.123E}. The interaction between dark matter and the ICM not only shapes the structure of galaxy clusters but also influences the evolution of galaxies. Studying the ICM-dark matter connection in galaxy clusters is of paramount importance for several reasons: 1) understanding dark matter \citep{2019A&A...628A..86B,2022A&A...662A.123E}, 2) constraining cosmological parameters \citep{2011ARA&A..49..409A,2024A&A...689A.298G}, 3) accounting for the cosmic baryonic budget \cite{2023A&A...675A.188A,2025A&A...702A.182R}, and 4) investigating feedback mechanisms \citep{2023MNRAS.518.2735I}. However, unveiling the secrets of this interplay requires innovative strategies to address the inherent challenges posed by the elusive nature of dark matter, the complex dynamics of the ICM, and the observational limitations such as projection effects.

Hydrodynamical simulations of galaxy clusters \citep{cui2018,2024MNRAS.533.2656B,2024A&A...687A.129L} offer a complementary approach to deciphering the complex interplay between dark matter and the ICM, effectively bridging gaps left by observational limitations. These simulations can model the intricate physical processes governing galaxy clusters, including gravitational interactions dominated by dark matter, thermal and non-thermal processes in the ICM, and feedback mechanisms from supernovae and active galactic nuclei (AGNs)  
 \citep{2015ApJ...813L..17R,2022MNRAS.514..977C,2024A&A...686A.157N}.
Moreover, they allow one to generate synthetic observations that mimic real telescope data, accounting for projection effects and instrumental noise. Such simulations not only help disentangle the three-dimensional structure of galaxy clusters from two-dimensional projections \citep{2023A&A...679A..51I}, they also serve as a benchmark for refining cosmological models \citep{2011MNRAS.417.2020S}.

Precise mass estimates of galaxy clusters are fundamental for calibrating X-ray scaling relations, such as the mass-luminosity and mass-temperature relations \citep{2009ApJ...692.1033V, 2010MNRAS.406.1759M}. As surveys such as eROSITA and the South Pole Telescope (SPT) push the boundaries of precision cluster cosmology \citep{2024A&A...689A.298G}, the integration of galaxy cluster scaling calibration techniques becomes crucial \citep{{lov22,kay22}}. The hydrostatic equilibrium method has long been the bedrock for estimating dark matter profiles in galaxy clusters using the X-ray and Sunyaev–Zel'dovich (SZ) data of the ICM \citep{2020A&A...634A.113A,2024EPJWC.29300050S}. 
However, this assumption may break down in clusters undergoing mergers or other dynamical processes, where deviations from equilibrium can lead to biased mass estimates. Non-thermal pressure contributions, arising from turbulent gas motions, cosmic rays, and additional physical processes, have been found, particularly through hydrodynamical simulations, to significantly contribute to the pressure support of the ICM and to introduce biases in hydrostatic mass estimates. \citet{2009ApJ...705.1129L} showed that the contribution of non-thermal pressure support increases with cluster radius and is particularly important in the outer regions of galaxy clusters, leading to an underestimate of the true cluster mass when hydrostatic equilibrium is assumed. Similarly, \citet{2014MNRAS.442..521S} investigated the evolution of non-thermal pressure support and demonstrated its dependence on cluster dynamical state and redshift, highlighting its importance for understanding hydrostatic mass bias. These findings are consistent with other simulation studies showing that non-thermal pressure support can produce significant deviations from hydrostatic equilibrium \citep{2013MNRAS.428.2366V,2024A&A...687A..58D}. 

Advanced mass-calibration methods, particularly weak-lensing and multi-wavelength approaches, provide independent constraints on cluster masses and help reduce uncertainties and systematic biases in cluster mass estimates. Weak-lensing measurements are especially valuable for calibrating hydrostatic mass bias, as they provide a mass estimate that is largely independent of the assumption of hydrostatic equilibrium. Comparisons between weak-lensing and hydrostatic mass measurements have therefore provided important observational constraints on the level of hydrostatic bias \citep{2014MNRAS.439....2V,2015MNRAS.449..685H,2020MNRAS.497.4684H}. A comprehensive review of weak-lensing techniques and their application to galaxy cluster mass measurements has been presented by \citet{2020A&ARv..28....7U}. Together with multi-wavelength observations, the advances in mass calibration are improving the reliability of cluster mass estimates and reducing the systematic uncertainties affecting cluster scaling relations \citep{2020MNRAS.492.3685H,2024PhRvD.110h3509B,2024A&A...682A.147M,2026arXiv260211989T}.

Machine learning techniques have notably advanced the precision of galaxy cluster mass measurements. \citet{Ntampaka} introduced support distribution machines to reconstruct dynamical cluster masses. Similarly, \citet{2019Ho,2020Ramanah,2021ApJ...908..204H,2022NatAs...6..936H} employed convolutional neural networks (CNNs) to infer cluster masses using relative line-of-sight velocities and projected radial distances of galaxy pairs. \citet{2019ApJ...876...82N} and \citet{2024A&A...682A.132K} developed CNN-based models to estimate galaxy cluster masses directly from X-ray photon data. \citet{2022arXiv220712337F} adopted a hybrid approach combining autoencoders and random forest regression, achieving an approximately 10\% scatter in reconstructing mass profiles and total cluster mass. For the first time, \citet{2022NatAs...6.1325D} used observational maps, specifically Planck $y$-maps, to directly estimate the mass,\footnote{The term M$_{\Delta}$ represents the total mass within the radius $\textrm{R}_{\Delta}$, defined as ${\rm M}_{\Delta}= 4/3 \pi \textrm{R}_{\Delta}^3 \, \Delta \, \rho_{\rm crit}(z)$, where $\rho_{\rm crit}(z)$ is the critical density at redshift $z$. Throughout the paper, we consider the overdensities of $\Delta=2500$, $500,$ and $200$.} M$_{500}$. In a related effort, \citet{2020ApJ...900..110G} developed a modified residual U-Net (mResUNet) to directly estimate cluster masses from simulated microwave-sky maps containing the thermal SZ signal together with Cosmic Microwave Background (CMB) emission, astrophysical foregrounds, and instrumental noise. The network was trained to learn the relation between the spatial information in these microwave observations and the cluster mass. \citet{2021ApJ...923...96G} subsequently extended this deep-learning framework to a two-stage approach in which one mResUNet reconstructs foreground- and noise-suppressed CMB maps, while a second network estimates cluster masses from the CMB gravitational-lensing signal. Similarly, utilising multi-wavelength maps, \citet{2024MNRAS.528.1517D} applied a U-Net architecture to predict the projected total mass density maps from hydrodynamical simulations. Recently, \citet{Iqbal2025} used graph neural networks to estimate masses.

In this work, we model the radial profiles of the ICM as sequential data, reflecting the fact that physical quantities such as gas density, temperature, pressure, and gas mass are intrinsically coupled along the radial direction. These properties are governed by the underlying gravitational potential and thermodynamic processes within galaxy clusters, meaning that variations at one radius are physically connected to conditions at neighbouring radii. This inherent ordering and interdependence make radial ICM profiles naturally suited to recurrent neural network (RNN) architectures, which are designed to capture sequential relationships and can flexibly accommodate variable-length inputs. We employed a gated recurrent unit (GRU; \citealt{cho-etal-2014-learning}) model to learn the radial dependencies within the ICM and to predict the underlying three-dimensional total matter mass profile, M(<R), defined as the mass enclosed within radius R.
By explicitly assuming the ordered structure of the ICM profiles, our approach enabled the model to learn complex non-linear mappings between spherically symmetric observable thermodynamic properties and the underlying true mass profiles. We trained the GRU on high-resolution hydrodynamical simulations from the GADGET-X and GIZMO-SIMBA suites within The Three Hundred Project \citep{cui2018}, which reproduce observed cluster scaling relations and pressure profiles reasonably well  \citep{2023MNRAS.523.1228L}. 

The structure of this paper is as follows: In Sect. 2, we describe the data and simulation setup. In Sect. 3, we detail the GRU model and training procedure. In Sect. 4, we present the results of our mass estimation method and compare it with traditional approaches. In Sect. 5, we apply our model to two observational X-ray samples of galaxy clusters, and in Sect. 6, we discuss the implications of our findings and future directions for research. Throughout this work, we assume a flat Lambda Cold Dark Matter ($\Lambda$CDM) model with $H_0 = 67.77$ km s$^{-1}$ Mpc$^{-1}$, $\Omega_m = 0.30$, and $\Omega_\Lambda = 0.70$. 
\section{Simulations}
\label{sec:simulations}
In this work, we use spherically averaged radial profiles of the gas and total matter (dark matter plus baryonic matter) distributions of galaxy clusters from the \thethree\ dataset \citep{cui2018}. In this section, we summarise the simulation setup (Section~\ref{sec:sub0}), the dynamical classification of clusters (Section~\ref{sec:sub1}), the derivation of hydrostatic masses (Section~\ref{sec:sub2}), and the definition of cluster sparsity used in this work (Section~\ref{sec:spar}).
\subsection{Simulated cluster sample}
\label{sec:sub0}
The cluster sample is drawn from \thethree,\, a suite of hydrodynamical simulations comprising 324 regions that encompass the most massive galaxy clusters selected from the MultiDark Planck 2 (MDPL2) cosmological N-body simulation. The initial conditions were generated by resimulating Lagrangian regions centred on halos identified at $z=0$, ensuring a mass-complete sample of cluster-scale objects. Each region is evolved using two independent hydrodynamical codes, GIZMO-SIMBA \citep{2022MNRAS.514..977C} and GADGET-X \citep{2016MNRAS.455.2110B}, which adopt the same underlying cosmology but differ in their hydrodynamical solvers and subgrid galaxy formation models. As we show later, this dual implementation allows for a direct assessment of systematic uncertainties associated with baryonic physics. Both the parent MDPL2 simulation and the resimulations assume a flat $\Lambda$CDM cosmology consistent with the Planck results \citep{2016A&A...594A..13P}. The resimulated regions include a comprehensive treatment of baryonic physics, including radiative cooling, a uniform UV background, star formation, supernova feedback, chemical enrichment, and active galactic nucleus (AGN) feedback \citep{2015ApJ...813L..17R}.

\begin{figure}
\centering
		\includegraphics[width=0.4\textwidth]{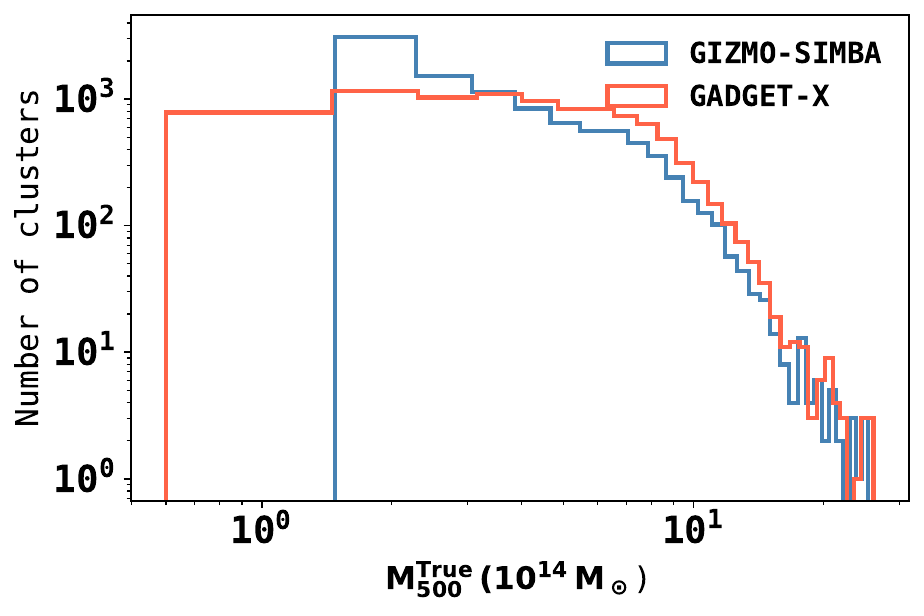}
		\caption{\footnotesize Distribution of galaxy clusters in \thethree\ as a function of true mass M$^{\mathrm{True}}_{500}$ for the GIZMO-SIMBA and GADGET-X simulation suites in the redshift range $0 \lesssim z \lesssim 1$, containing 9997 and 9590 clusters, respectively. }
		\label{figSec1}
\end{figure}

In this work, we used 19 and 31 snapshots from the GIZMO-SIMBA and GADGET-X simulation suites, respectively, spanning the redshift range $0 \lesssim z \lesssim 1$. The GIZMO-SIMBA sample contains 9997 clusters with masses in the range $1.5 \times 10^{14}\,\mathrm{M}_\odot \lesssim \rm{M}_{500} \lesssim 2.5 \times 10^{15}\,\mathrm{M}_\odot$, while the GADGET-X sample includes 9590 clusters with masses $6.0 \times 10^{13}\,\mathrm{M}_\odot \lesssim \rm{M}_{500} \lesssim 2.6 \times 10^{15}\,\mathrm{M}_\odot$. The cluster samples include the main progenitors of the present-day ($z=0$) massive clusters identified through the simulation merger trees as well as other companion objects present in the resimulated region.
For each simulated cluster, we extracted spherically averaged three-dimensional profiles of the ICM, including gas density (n$_g$) and temperature ($\rm{T}_g$), gas pressure (P$_{\rm g}$) and integrated gas mass M$_{\rm g}$, and derived the corresponding spherically averaged total integrated mass profiles, M(<R). Profiles were sampled on a logarithmic radial grid of 40 points spanning $0.05\,\rm{R}_{500}$ to $2\,\rm{R}_{500}$. However, the inner radial boundary was defined by a convergence criterion requiring a minimum of 100 gas particles enclosing it.  Figure~\ref{figSec1} shows the distribution of galaxy clusters in \thethree\ as a function of the true halo mass, M$^{\mathrm{True}}_{500}$, for the GIZMO-SIMBA and GADGET-X simulations. Appendix \ref{app00} gives the detailed comparison of temperature and density profiles of the GIZMO-SIMBA and GADGET-X simulations from \thethree\, project. 
\subsection{Dynamical state of the simulated clusters}
\label{sec:sub1}
To investigate the impact of cluster dynamics on the performance of the GRU models, we estimated the dynamical states of all clusters in the GIZMO-SIMBA simulation suite. Clusters were classified according to their dynamical state at overdensities $\Delta = 500$ and $200$, using the substructure mass fraction, $f_{s}=\textrm{M}_{\rm s}/\textrm{M}$, and the offset between the central density peak and the centre of mass, $\Delta_{r}=|r_{\rm peak} - r_{\rm cm}|/\textrm{R}$ \citep{Rasia2013}. Relaxed clusters are defined by both $f_{s}$ and $\Delta_{r}$ being below 0.1 \citep[e.g.][]{Cui2017,Cialone2018,DeLuca2021}. The computation of the dynamical indicators was extended to both R$_{500}$ and R$_{200}$.

Similar to \citet{Rasia2013}, we combined these indicators into a single dynamical-state parameter ($\chi$), defined as
\begin{eqnarray}
\chi_{_\delta}
&=&
\frac{1}{2}
\left[
\frac{\Delta_{r}-\widetilde{\Delta}_{r}}
{Q_{75}(\Delta_{r})-Q_{25}(\Delta_{r})}
+
\frac{f_{s}-\widetilde{f}_{s}}
{Q_{75}(f_{s})-Q_{25}(f_{s})}
\right],
\label{eq:chi}
\end{eqnarray}
where $\widetilde{\Delta}_{r}$ and $\widetilde{f}_{s}$ denote the median values of the corresponding distributions, and $Q_{25}$ and $Q_{75}$ are the first and third quartiles, respectively. This normalisation places both indicators on comparable scales while reducing sensitivity to outliers. The resulting parameters, $\chi_{_{500}}$ and $\chi_{_{200}}$, provide continuous measures of the dynamical state of clusters within R$_{500}$ and R$_{200}$, respectively. Lower (negative) values correspond to dynamically relaxed systems, whereas higher (positive) values indicate progressively more disturbed clusters. 
\subsection{Hydrostatic mass profile}
~\label{sec:sub2}
The hydrostatic mass profiles were computed using the hydrostatic equilibrium equation, which links the pressure gradient to the enclosed gravitational mass as 
\begin{equation}
\textrm{M}^{\rm HSE}(\textrm{<R}) =-  \frac{\textrm{R}^2}{G} \frac{d\textrm{P}_{g}}{d\textrm{R}} \frac{1}{{\rm n}_{g}}
\label{H:mass},
\end{equation}
where $G$ is the gravitational constant. Before using the above equation, the pressure profiles were first smoothed in logarithmic space using locally linear weighted regression to compute the pressure derivatives \citep{Cappellari2013b}. This approach employs a tri-cube weighting function, where the weight of each point decreases with distance from the local radial bin, and a smoothing parameter $f$ which controls the fraction of neighbouring points included in the local fit; larger $f$ yields smoother profiles. We adopt a modest smoothing and consider $f$ as a dynamic parameter, which changes from 0.10 to 0.20 from outer to inner regions. At each radius, the local linear fit minimises a weighted least-squares error, constrained such that the local slope is non-positive, producing a smoother estimate of the pressure and its derivative. The resulting logarithmic derivative was then converted back to linear space and combined with the gas density to compute the hydrostatic mass.  After performing the fit, we observed that a few points, particularly at small radii (where the profiles are relatively more irregular), resulted in best-fitting slope values of zero, which corresponds to the upper bound of the fit. This indicates that at these points, the fit would actually favoured positive slopes, which is inconsistent with the hydrostatic equilibrium assumption. To address this, we removed these problematic points and recomputed the fit.
\begin{figure}
\centering
		\includegraphics[width=0.4\textwidth]{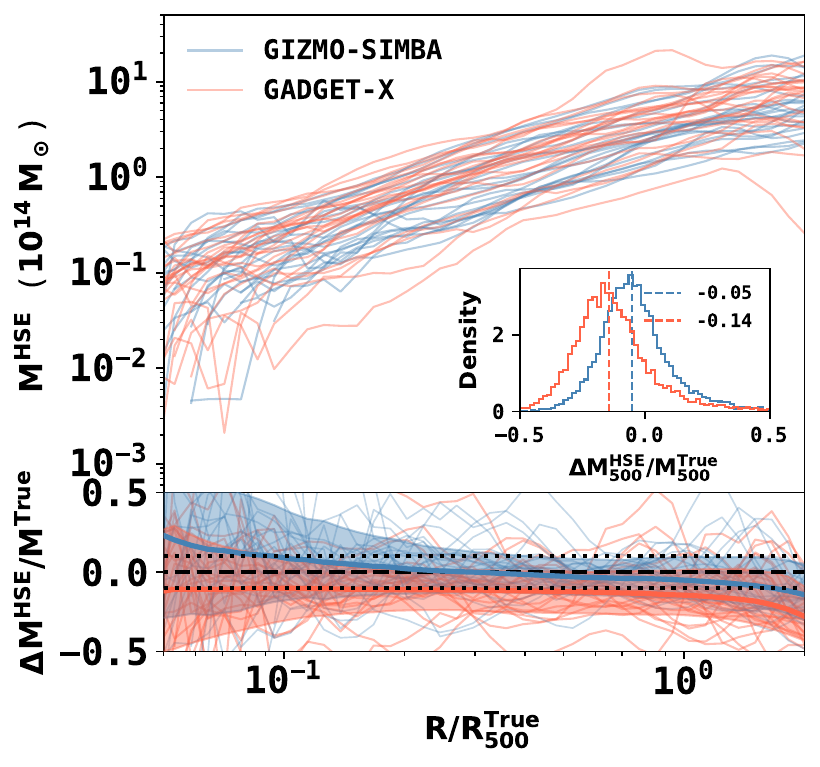}
		\caption{\footnotesize Hydrostatic mass profiles of 25 randomly selected clusters from each of the GIZMO-SIMBA and GADGET-X simulations. The lower panel shows the corresponding fractional residuals relative to the true mass profiles. For the full cluster sample in each simulation, the thick solid line denotes the median residual profile, while the shaded region represents the spread between the  $84^{\rm th}-16^{\rm th}$ percentile range. The inset panel displays the distribution of hydrostatic mass bias measured at R$_{500}$, yielding median offsets of $-0.05$ for GIZMO-SIMBA and $-0.14$ for GADGET-X. Horizontal dotted black lines indicate the $\pm10\%$ bias levels, and the dashed black line represents zero bias.}
		\label{fig1}
\end{figure}
\begin{figure*}
    \centering
    \includegraphics[width=0.80\textwidth]{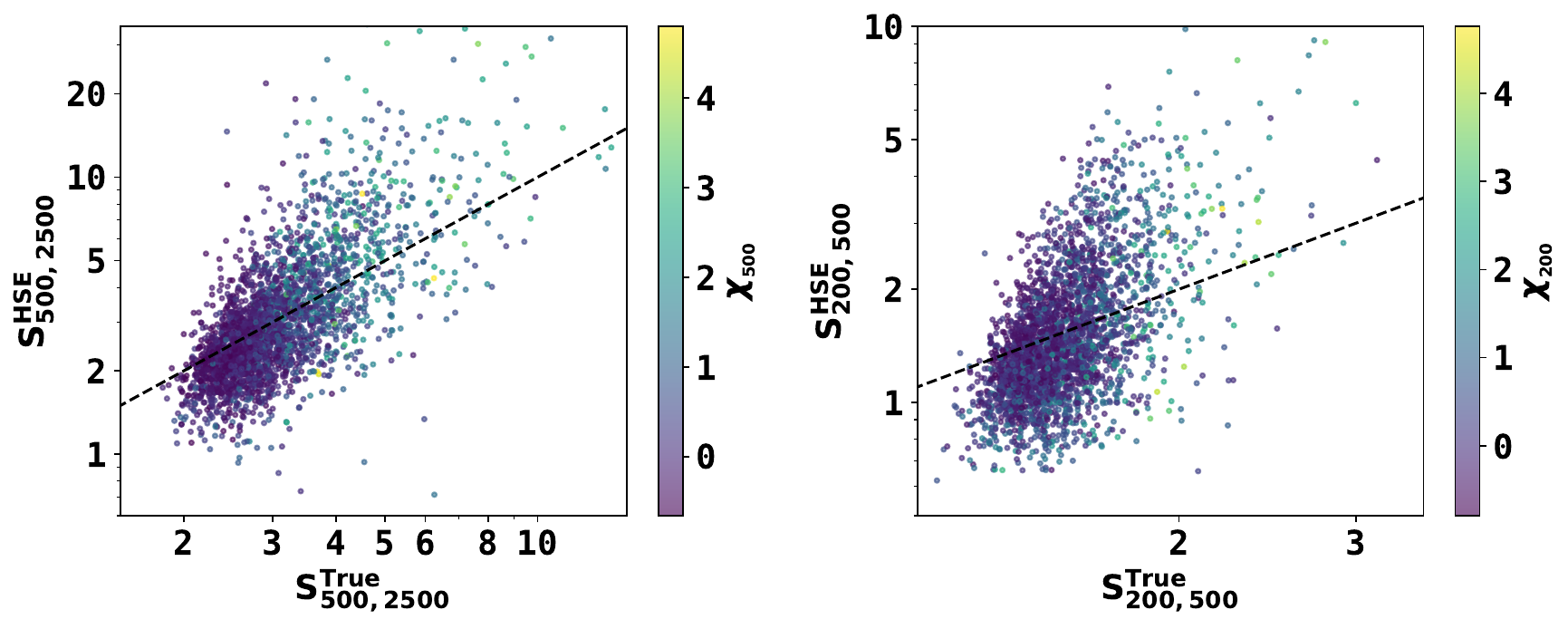}
\caption{\footnotesize Cluster sparsity derived from hydrostatic mass profiles, S$^{\rm HSE}$, compared with the corresponding true sparsity values for the overdensity pairs S$^{\rm True}_{500,2500}$ (left) and S$^{\rm True}_{200,500}$ (right) based on a randomly selected sample of 3000 galaxy clusters from the GIZMO-SIMBA simulations.  
 Points are colour-coded by the dynamical state parameter $\chi$, with lower sparsity values associated with more relaxed systems (smaller $\chi$), while higher sparsity values correspond to dynamically disturbed clusters. Values of S$^{\rm HSE}_{200,500}<1$ arise primarily from strongly non-monotonic or non-increasing enclosed mass profiles, M$(<{\rm R})$. The dashed black line represents the one-to-one relation.
}
    \label{fig:sparsity}
\end{figure*}

Figure~\ref{fig1} presents the hydrostatic mass profiles of 25 randomly selected clusters drawn from each of the GIZMO-SIMBA and GADGET-X simulations. The corresponding fractional residuals, calculated with respect to the hydrostatic mass profiles derived from the full cluster sample of each simulation, are also shown. The hydrostatic mass residuals exhibit distinct radial trends in the two simulation suites. In the central regions (R$ \lesssim 0.1\,{\rm R}_{500}$), GIZMO-SIMBA systematically overestimates the true mass, whereas GADGET-X shows no evidence of such a central excess. This difference likely reflects variations in the treatment of gas dynamics and baryonic physics, which affect the pressure gradients used in the hydrostatic mass reconstruction. At intermediate radii, GADGET-X exhibits a nearly constant hydrostatic mass bias, with median fractional residuals ($\Delta {\rm M}_{500}/{\rm M}^{\rm True}_{500}$) of approximately $-10\%$, while the GIZMO-SIMBA residuals slightly decrease with radius but remain closer to zero. Beyond R$_{500}$, the median fractional residuals in both simulations become progressively more negative, indicating an increasing hydrostatic mass underestimate towards larger radii. This trend is stronger in GADGET-X than in GIZMO-SIMBA, leading to a larger hydrostatic mass bias in the cluster outskirts.

We find the median hydrostatic mass bias, defined as $(1-b) \equiv {\rm M}^{\rm HSE}/{\rm M}^{\rm True}$, to be $0.95_{-0.12}^{+0.15}$ for the GIZMO-SIMBA sample and $0.86_{-0.13}^{+0.15}$ for the GADGET-X sample at R$_{500}^{\rm True}$, where the quoted uncertainties correspond to the  $84^{\rm th}-16^{\rm th}$ percentile range. The scatter\footnote{We quantify scatter, $\sigma_{\rm scatter}$, as the standard deviation of the logarithmic residuals.}, evaluated at R$_{500}^{\rm True}$, is $\sigma_{\rm scatter}=0.067$ dex and $0.081$ dex for GIZMO-SIMBA and GADGET-X samples, respectively. Figure~\ref{figA2} in Appendix~\ref{app_pressure_bias} shows the pressure profiles of the 100 clusters with the lowest and highest absolute hydrostatic mass bias in our sample, together with the corresponding smooth profiles reconstructed using our local fitting scheme. The reconstructed pressure profiles are typically accurate to within $\sim 10\%$ of the true profiles over most of the cluster region and show no evidence of a systematic bias. 
\subsection{Halo sparsity}
\label{sec:spar}

The sparsity of a galaxy cluster is defined as the ratio of the cluster mass enclosed within two spherical overdensities \citep{2014MNRAS.437.2328B,2018ApJ...862...40C}, and provides a simple probe of the halo density profile. In this work, we define the sparsity such that $\Delta_1 > \Delta_2$:
\begin{equation}
{\rm S}_{\Delta_1, \Delta_2} = \frac{{\rm M}_{\Delta_1}}{{\rm M}_{\Delta_2}}.
\end{equation}
Halo sparsity is a valuable cosmological probe, as comparing the mass distribution of halos at different overdensities allows for constraints on parameters such as the matter density $\Omega_m$ and the amplitude of matter fluctuations $\sigma_8$ \citep{2025A&A...697A..33C}.  Figure~\ref{fig:sparsity} presents the cluster sparsity for randomly selected 3000 galaxy clusters derived from GIZMO-SIMBA simulations, estimated from hydrostatic equilibrium mass profiles (S$^{\rm HSE}$), compared with the true values from simulations for the two sparsity definitions S$^{\rm HSE}_{500,2500}$ and S$^{\rm HSE}_{200,500}$. We find that the lower sparsity values are predominantly associated with clusters exhibiting smaller values of the dynamical state parameter $\chi$, indicative of more dynamically relaxed systems, whereas higher sparsity values are generally associated with dynamically disturbed clusters. 

This behaviour can be understood in terms of the underlying mass distribution. Relaxed clusters, which are dynamically settled and exhibit minimal substructure and small centroid offsets, tend to have more centrally concentrated mass profiles. Consequently, for a given overdensity, a larger fraction of the halo mass is contained within smaller radii, meaning that ${\rm M}_{2500}$ constitutes a relatively larger fraction of ${\rm M}_{500}$. This leads to lower sparsity values. In contrast, merging or disturbed clusters generally exhibit shallower and less regular mass profiles due to ongoing dynamical activity, unrelaxed substructures, and larger centroid shifts. As a result, the inner mass ${\rm M}_{2500}$ is reduced relative to the total mass ${\rm M}_{500}$, yielding higher sparsity values. These trends highlight the sensitivity of halo sparsity to both the dynamical state of clusters and the adopted mass estimation method. For the full GIZMO-SIMBA sample, we find scatter to be 0.22 dex and 0.13 dex for ${\rm S}^{\rm HSE}_{500,2500}$ and ${\rm S}^{\rm HSE}_{200,500}$, respectively. Similarly, for the GADGET-X, we find scatter to 0.19 dex and 0.17 dex for ${\rm S}^{\rm HSE}_{500,2500}$ and ${\rm S}^{\rm HSE}_{200,500}$, respectively.

We further note that, for a subset of clusters, the hydrostatic mass estimates at R$_{200}$, M$^{\rm HSE}_{200}$, are unphysically smaller than M$^{\rm HSE}_{500}$. This behaviour originates from a higher level of non-monotonic hydrostatic mass profiles beyond the true R$_{500}$, likely caused by departures from hydrostatic equilibrium and increased dynamical activity in the cluster outskirts. These systems can be identified in Fig.~\ref{fig:sparsity} by sparsity values S$^{\rm HSE}_{200,500}<1$ and are also predominantly associated with large values of the dynamical state parameter $\chi$, indicative of disturbed clusters.
\section{GRU model for predicting cumulative total mass profiles}
In this section, we present a detailed description of the GRU-based model developed to predict the true radial three-dimensional total mass profiles of galaxy clusters (dark and baryonic matter) from three-dimensional radial ICM observables. The GRUs are a class of RNNs that have proven highly effective for modelling sequential and ordered data, and have seen increasing use in astrophysical and cosmological applications \citep{2017ApJ...837L..28C,2024A&A...682A.177S}.  We begin by motivating the treatment of ICM profiles as sequential data, followed by a brief overview of GRU fundamentals. We then describe the specific GRU-based architecture used in this work, highlighting how it incorporates variable-length radial profiles, redshift conditioning, and uncertainty-aware predictions.
\subsection{ICM as sequential data}
We model the radial profiles of ICM as sequences and employ a GRU network to map these sequences to total cumulative mass profiles, M(<R), of galaxy clusters. 
The sequential nature of ICM profiles can be argued from the strong physical coupling between neighbouring radial bins. As one moves outwards from the cluster centre, the gas density typically decreases, while temperature and pressure follow characteristic radial trends shaped by gravity, radiative processes, and gas dynamics. These properties do not evolve independently; rather, they exhibit correlated, non-linear dependencies across radius. Formally, we define the input feature vector at radial bin $i$ as
\begin{equation}
\mathbf{x}_i = \begin{bmatrix}
\textrm{T}_{g,i},  {\rm n}_{g,i}, \textrm{P}_{g,i}, \textrm{M}_{g,i}, \textrm{R}_i 
\end{bmatrix},
\end{equation}
where $\textrm{T}_{g,i}$, ${\rm n}_{g,i}$, $\textrm{P}_{g,i}$, $\textrm{M}_{g,i}$, and $\textrm{R}_i$ denote the temperature, gas density, pressure, enclosed gas mass, and radius at that bin, respectively. The full ICM profile for a cluster is thus represented as an ordered sequence
$\{\mathbf{x}_1, \mathbf{x}_2, \dots, \mathbf{x}_N\}$, where $N$ is the number of bins and could vary from cluster to cluster. The GRU takes this sequence as input and predicts the corresponding total cumulative mass profile as a function of radius, M(<R). 

\subsection{GRU overview}
The GRUs, introduced by \citet{cho-etal-2014-learning}, are designed to overcome key limitations of traditional RNNs, most notably the vanishing gradient problem. GRUs achieve this through gating mechanisms that regulate the flow of information across time steps, enabling the network to retain long-term dependencies while remaining computationally efficient. Compared to long short-term memory networks \citep{10.1162/neco.1997.9.8.1735}, GRUs employ a simpler structure with fewer parameters, often achieving comparable performance at lower computational cost. Compared to standard neural networks, as other RNNs, GRUs naturally process multiple dimensional sequential data without requiring fixed-size vectorisation.

At each sequence step $t$, a GRU processes the current input $\mathbf{x}_t$ and the previous hidden state $\mathbf{h}_{t-1}$ to produce an updated hidden state $\mathbf{h}_t$ governed by two gating mechanisms: the update gate $\mathbf{g}_{u,t}$ and the reset gate $\mathbf{g}_{r,t}$. The update gate controls how much information from the previous hidden state is retained, while the reset gate determines how much of the past information is used when constructing the hidden state. The GRU operations are
\begin{eqnarray}
\mathbf{g}_{u,t} &=& \sigma\!\left(
W_u [\mathbf{h}_{t-1}, \mathbf{x}_t] + \mathbf{b}_u
\right)\nonumber\\
\mathbf{g}_{r,t} &=& \sigma\!\left(
W_r [\mathbf{h}_{t-1}, \mathbf{x}_t] + \mathbf{b}_r
\right)\nonumber\\
\tilde{\mathbf{h}}_t &=& \tanh\!\left(
W_h [\mathbf{g}_{r,t} \odot \mathbf{h}_{t-1}, \mathbf{x}_t]
+ \mathbf{b}_h
\right)\nonumber\\
\mathbf{h}_t &=&
(1-\mathbf{g}_{u,t}) \odot \mathbf{h}_{t-1}
+\mathbf{g}_{u,t} \odot \tilde{\mathbf{h}}_t,
\end{eqnarray}
where $\sigma(\cdot)$ denotes the sigmoid activation function, $\odot$ indicates element-wise multiplication, and $[\cdot,\cdot]$ represents vector concatenation. The learnable parameters $W_u$, $W_r$, and $W_h$ are weight matrices of dimension $(d_h, d_h + d_x)$, with corresponding bias vectors $\mathbf{b}_u$, $\mathbf{b}_r$, and $\mathbf{b}_h$, where $d_x$ and $d_h$ denote the input and hidden state dimensions, respectively.
\subsection{GRU architecture for cumulative total mass prediction}
\label{subsec:gru_architecture}
We employed a multi-layer GRU network to model the sequential nature of ICM radial profiles of variable length, enabling prediction of the total enclosed mass profile at the corresponding ICM radii. 
Given an input sequence $\mathbf{X} \in \mathbb{R}^{N \times 5}$, where $N$ is the number of radial bins and the five features correspond to values of pressure, gas mass, radius, density, and temperature, the GRU produces a sequence of hidden states $\mathbf{H} \in \mathbb{R}^{N \times d_h}$ (or $\mathbb{R}^{N \times 2d_h}$ in the bidirectional case). Furthermore, the GRU outputs are explicitly conditioned on the cluster redshift $z$. The redshift is projected into the GRU's latent space via a learnable linear transformation, repeated across all radial bins, and concatenated with the GRU features at each bin. This conditioning mechanism allows the network to combine information from the radial profiles with global redshift context, thereby accounting for systematic evolutionary effects (if any) in the ICM--mass relation.

At each radius R$_i$, the final network layer outputs two parameters: the mean $\mu_i$' and the logarithmic\footnote{Throughout this work, $\log$ without a specified base denotes the natural logarithm, i.e. base $e$.} variance `$\log\sigma_i^2$' for the log-normal distribution of the total mass, i.e. $\log_{10} {\rm M}(<{\rm R}_i)$ as
\begin{equation}
\log_{10} {\rm M}(<{\rm R}_i) \equiv \mathcal{N}(\mu_i, \sigma_i^2).
\end{equation}
To provide more robust, radius-dependent uncertainty estimates, we trained an ensemble of 50 independent GRU models with different initialisations. Predictions were combined using weights $w_i$ inversely proportional to each model's training loss. For an ensemble of $K=50$ models, the ensemble mean `$\mu_{\text{ens}}$' and variance `$\sigma_{\text{ens}}^2$' at each radius were computed as
\begin{eqnarray}
\mu_{\text{ens}}({\rm R}) &=& \sum_{i=1}^{K} w_i \times \mu_i({\rm R}) \nonumber\\
\sigma_{\text{ens}}^2({\rm R}) &=&
\sum_{i=1}^{K} w_i \times \sigma_i^2({\rm R}) +
\sum_{i=1}^{K} w_i \times \left[\mu_i ({\rm R}) - \mu_{\text{ens}}({\rm R})\right]^2.
\label{eq:ensemble_stats}
\end{eqnarray}
The integrated mass profile in linear space and its $1\sigma$ uncertainty are then expressed compactly as  
\begin{equation}
{\rm M}^{\rm GRU}(\rm R) = 10^{\mu_{\text{ens}}(\rm R) \pm \sigma_{\text{ens}}(\rm R)}
\label{eq:confidence_interval},
\end{equation}
where the central value is $10^{\mu_{\text{ens}}(\rm R)}$ and the $1\sigma$ confidence interval in linear space is given by $\left[10^{\mu_{\text{ens}}(\rm R) - \sigma_{\text{ens}}(\rm R)},\; 10^{\mu_{\text{ens}}(\rm R) + \sigma_{\text{ens}}(\rm R)}\right]$.
\subsection{Training and testing samples}
To avoid data leakage between training and testing, we first randomly split the 324 independent zoom-in regions into 274 training regions and 50 test regions. All clusters from the training regions are used for model training, while those from the test regions are reserved exclusively for final evaluation. This split yields approximately 16495 galaxy clusters in the training sample and about 3092 in the test sample, both covering a redshift range of \(0 \leq z \leq 1\). Furthermore, radial sampling and coverage are strongly affected by observational constraints \citep[e.g.][]{2024A&A...688A.219C}. To account for this, we perform data augmentation on both the training and testing samples. Starting from the original high-resolution profiles defined on 40 radial grid points over the range \([0.05, 2]\, \mathrm{R}_{500}\), we generate degraded versions by randomly reducing the number of radial bins (with a minimum of 7 and a maximum of 40) and by varying both the inner and outer radial limits. The innermost radius is drawn from \([0.05, 0.3]\, \mathrm{R}_{500}\), while the outermost radius is sampled from \([0.75, 2]\, \mathrm{R}_{500}\). This procedure produces profiles with diverse radial extents, thereby better reflecting the quality, incompleteness, and heterogeneity of real observational data. 
\subsection{Model training and loss function}
\label{subsec:training}
All input quantities were normalised to improve numerical stability and accelerate convergence. Pressure, gas mass, radius, and total mass were divided by reference scales of $10^{-10}\textrm{erg}\, \textrm{cm}^{-3}$, $10^{14}\textrm{M}_{\odot}$, $1\textrm{Mpc}$, and $10^{15}\textrm{M}_{\odot}$, respectively, bringing their maximum values near unity. Similary, ICM density (in $\textrm{cm}^{-3}$) and temperature (in keV) profiles, instead, were not rescaled, as their natural numerical ranges were already suitable for stable gradient computation. All quantities were subsequently transformed to base-10 logarithmic space to better capture the multiplicative power-law relationships inherent to astrophysical scaling relations.

The model was trained to minimise the Gaussian negative log-likelihood (NLL) loss, which naturally accommodates the probabilistic output of mean ($\mu_i$) and log-variance ($\log\sigma_i^2$) at each radial bin. Given the target true cluster profiles, $y_i = \log_{10} {\rm M
}_{\text{true}}(<{\rm R}_i)$, with $N$ non-padded radial bins, the loss is computed as
\begin{equation}
\mathcal{L}_{\text{NLL}} = \frac{1}{\rm N} \sum_{i=1}^{N} \left[ \frac{1}{2} \log\sigma_i^2 + \frac{(y_i - \mu_i)^2}{2\sigma_i^2} \right].
\label{eq:nll_loss}
\end{equation}
The variance is parameterised in logarithmic space to ensure positivity during optimisation, thus improving numerical stability. This formulation enables joint learning of accurate mass predictions (via the squared error term) and appropriate, radius-dependent uncertainty estimates (via the variance term).

Hyperparameter optimisation was performed using \texttt{Optuna} \citep{akiba2019optuna}, which employs a Tree-structured Parzen Estimator algorithm \citep{2023arXiv230411127W} to efficiently search the parameter space by minimising training loss. Packed sequence processing was employed to enable efficient batched training on variable-length profiles while preventing padded bins from influencing the learned representations.
The final architecture employed the following optimal hyperparameters: Adam optimiser with a weight decay of $1\times10^{-3}$ and a learning rate of $7 \times 10^{-4}$, hidden dimension 120, dropout rate of 0.1, batch size of 64, and 2 bidirectional GRU layers. Furthermore, the learning rate was dynamically adjusted during training using a \texttt{ReduceLROnPlateau} scheduler, which reduced the rate by a factor of 0.5 when the validation loss plateaued for 10 consecutive epochs. Training typically converged within 400 epochs, after which no significant improvement in validation loss was observed. Tables~\ref{tab:hyperparameters} and \ref{tab:gru_architecture} in Appendix~\ref{app000} summarise the hyperparameters and provide a detailed overview of the GRU architecture, respectively.
\begin{figure*}
    \centering
    \begin{minipage}{0.4\linewidth}
        \includegraphics[width=\textwidth]{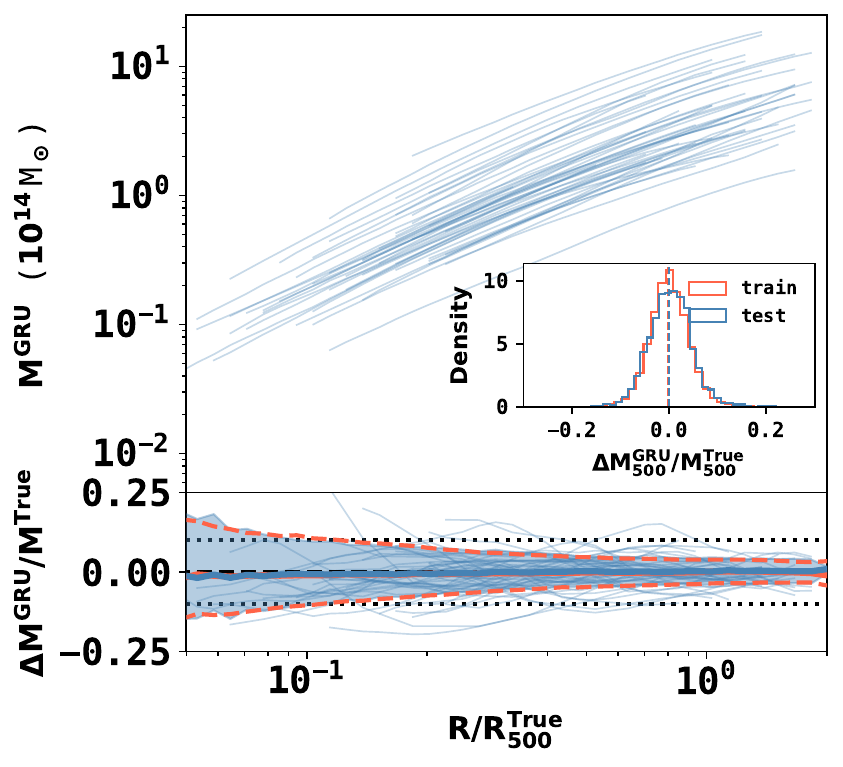} %
    \end{minipage}
    \begin{minipage}{0.4\linewidth}
        \includegraphics[width=\textwidth]{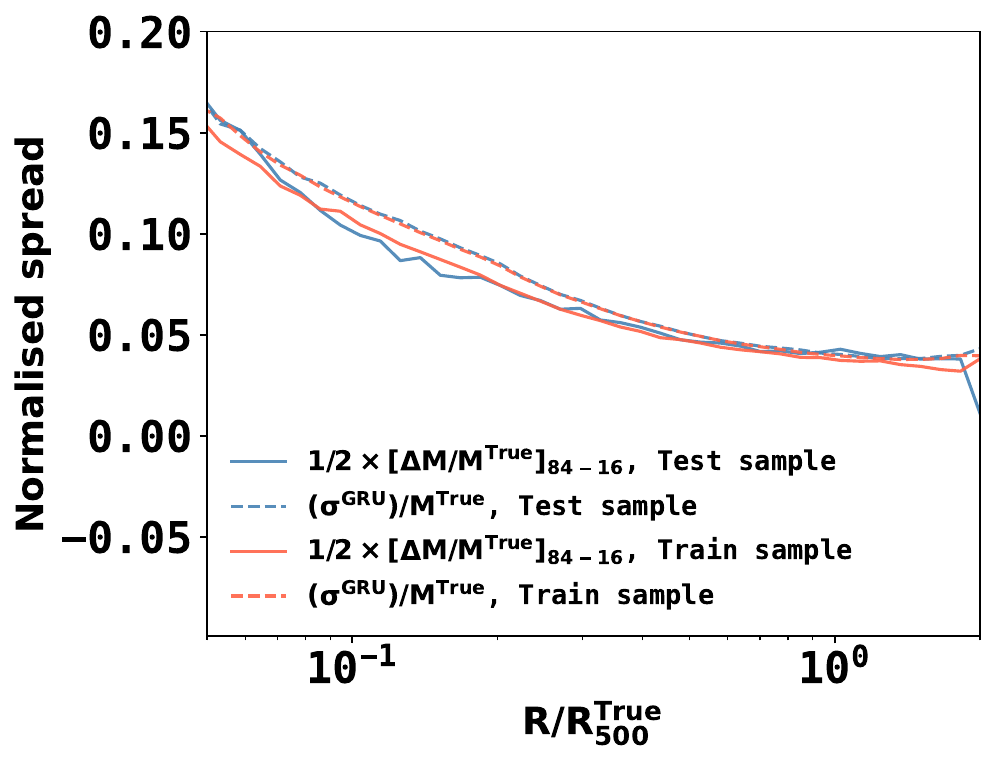}
    \end{minipage}
    \caption{\footnotesize GRU mass profile reconstructions and uncertainty estimates as a function of normalised radius.
The left panel shows the GRU mass profile reconstructions for 50 randomly selected clusters from the test sample, while the lower sub-panel displays the corresponding fractional residuals. The blue line and blue shaded region represent, respectively, the median and the $84^{\rm th}-16^{\rm th}$ percentile range of the fractional residuals for the full test sample. Similarly, the red line and the region enclosed by dashed red lines indicate, respectively, the median and the $84^{\rm th}-16^{\rm th}$ percentile range for the training sample. The horizontal dotted black lines mark the $\pm10\%$ fractional residual level, and the horizontal dashed black line denotes zero residual. The inset shows the distribution of fractional residuals in M$_{500}$  for the training and test samples. The right panel shows the radial dependence of the model uncertainty. The $(84^{\rm th}-16^{\rm th})/2$ percentile spread of the fractional mass residuals is shown as a function of radius for both the training (solid red line) and test samples (solid blue line). The corresponding GRU-predicted 1$\sigma$ fractional uncertainties are plotted as dashed red and blue lines, respectively. The close agreement between the measured residual spread and the predicted uncertainty, as well as the consistency between the training and test samples, demonstrates that the GRU provides well-calibrated and stable uncertainty estimates across the full radial range.} 
    \label{figR1} 
\end{figure*}

\begin{figure}
\centering
\includegraphics[width=0.45\textwidth]{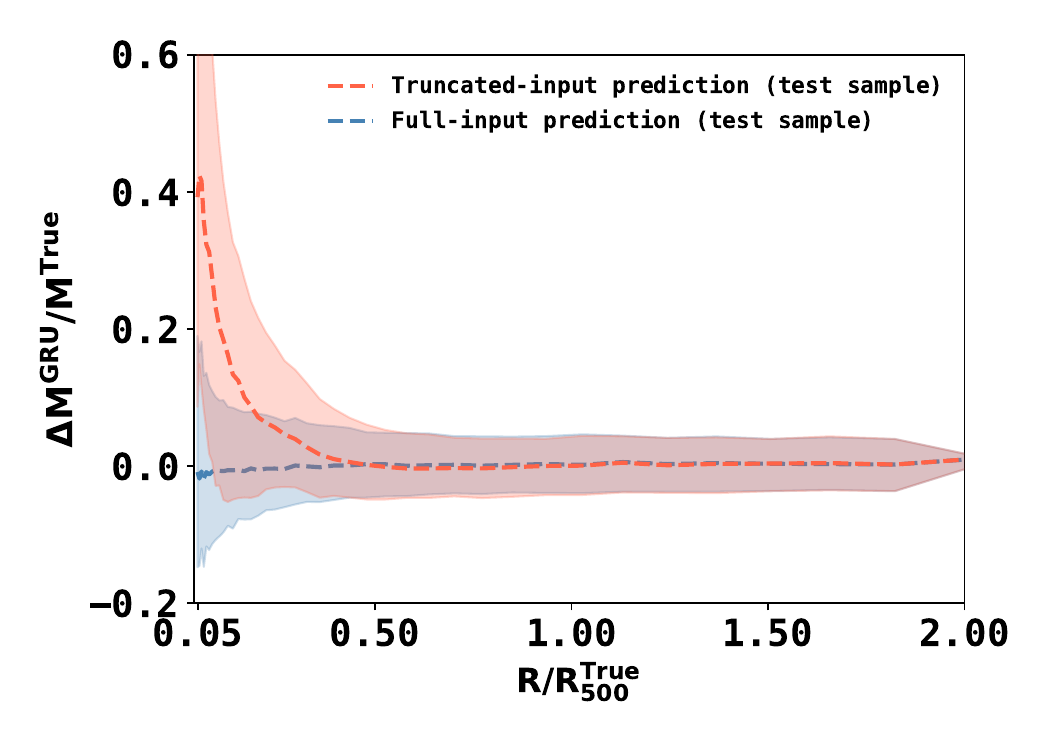}
\caption{\footnotesize Median fractional residual and $84^{\rm th}$--$16^{\rm th}$ percentile range (shaded bands) as a function of normalised radius for the test sample. The figure enables comparison of the full-input prediction with the truncated-input prediction. The full-input curve uses the entire radial profile to predict $\mathrm{M}(<\rm R)$; each point on the truncated-input
curve is a separate model evaluation using only the data interior to that radius. The truncated-input prediction is systematically biased high at small radii, reaching a median fractional residual of $\sim 40\%$ in the central region ($\mathrm{R} \lesssim 0.1\, \mathrm{R}_{500}$) and decaying to zero by $\mathrm{R} \sim 0.5\, \mathrm{R}_{500}$, at which point the truncated- and full-input curves
converge.}
\label{figR1_new}
\end{figure}

\begin{figure*}
\centering
\begin{minipage}{0.4\linewidth}
 \includegraphics[width=\textwidth]{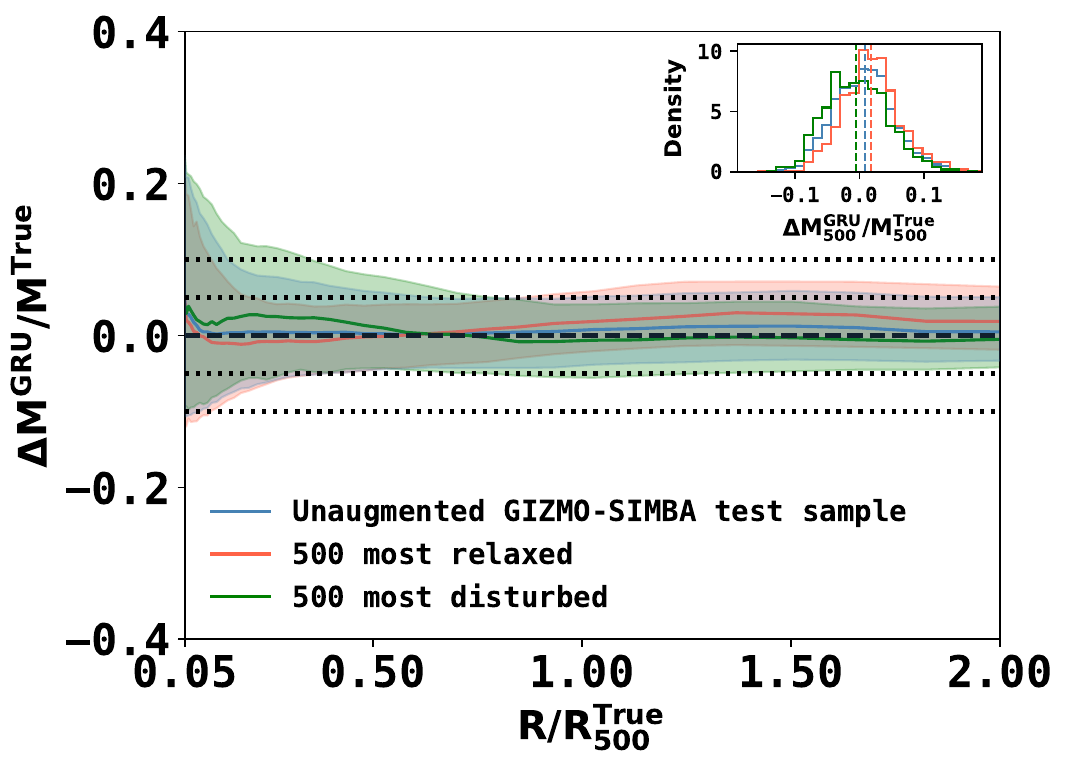} %
\end{minipage}
\begin{minipage}{0.4\linewidth}
\includegraphics[width=\textwidth]{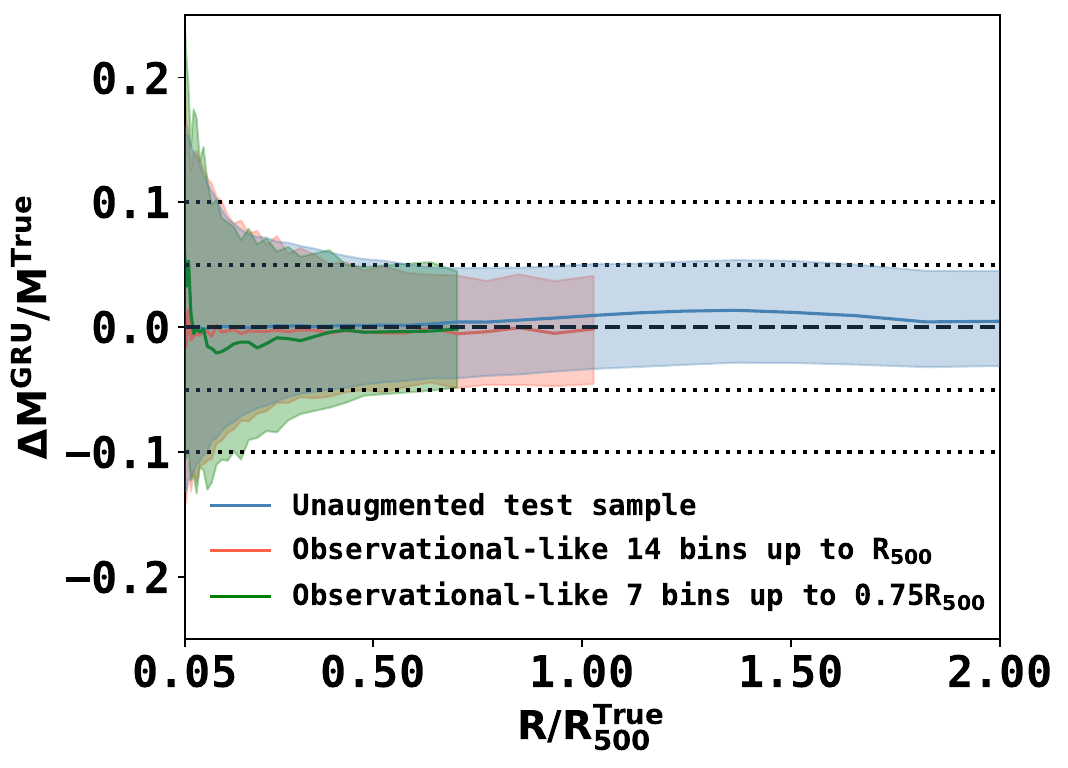}
\end{minipage}
\caption{Median fractional residuals of the mass profiles reconstructed with the GRU model for GIZMO-SIMBA test-sample clusters as a function of normalised radius. In both panels, the solid lines show the median residual profiles, while the shaded regions represent the $84^{\rm th}-16^{\rm th}$ percentile range. Horizontal black dotted lines indicate the $\pm 5\%$ and $\pm 10\%$ residual levels, and the dashed line marks zero residual. Left panel: Residuals for the full unaugmented GIZMO-SIMBA test sample comprising 1610 galaxy clusters together with the 500 most relaxed and 500 most disturbed clusters selected from the same sample. The inset shows the distribution of fractional residuals in M$_{500}$ for the three corresponding profile sets. Right panel: Comparison between reconstructions obtained from the original unaugmented profiles and from observation-like binned profiles, using either 14 logarithmically spaced radial bins out to R$_{500}$ or seven bins out to 0.75\,R$_{500}$. Despite the reduced radial resolution, the model maintains a comparable level of accuracy across all cases. 
}
    \label{figR2} 
\end{figure*}
\section{Model evaluation}
In this section, we assess the ability of our model to predict galaxy cluster mass profiles. We evaluate its robustness under realistic observational limitations, investigate how its predictions depend on the dynamical state of clusters, and benchmark its performance against the conventional hydrostatic mass estimation approach. We further examine the model’s generalisability across different simulation suites. In addition, we test alternative GRU configurations with varying input feature dimensionalities. Finally, we employ interpretability techniques to identify which radial thermal profile features most strongly influence the model’s mass predictions. Unless otherwise specified, all results are reported relative to our fiducial five-feature model.
\subsection{Performance with respect to the training and testing sample}
Model predictions are quantitatively evaluated against the ground-truth mass profiles obtained from simulations. The left panel of Fig.~\ref{figR1} presents the reconstructed total mass profiles obtained with the GRU based model for 50 randomly selected clusters from the testing sample, compared against the true mass profiles from hydrodynamic simulations. Also shown are the median fractional residuals for both the training and test samples. We find the model successfully recovers the overall shape of the cluster mass profiles, with no systematic bias, across the full radial range. The distribution of residuals, shown alongside the profiles, is sharply peaked around zero, further confirming the absence of systematic over- or underestimation of cluster mass profiles. The strong overlap between the training and test residual distributions suggests that the model does not suffer from overfitting. For both the training and testing samples, we find the median dispersion in the fractional residuals to be approximately 17\% at 0.05\,R$_{500}$. This dispersion decreases with radius and becomes nearly constant beyond 0.3\,R$_{500}$, stabilising at approximately $5\%$. The larger scatter in the central regions can be interpreted as a consequence of baryonic feedback processes, which reduce the correspondence between the ICM profiles and the underlying dark matter profiles. Overall, the scatter in M$_{500}$ across all clusters was $\sigma_{\rm scatter}=0.019$ dex for the test sample and $0.018$ dex for the training sample. 

To further quantify the radial dependence of the model performance, the right panel of Fig.~\ref{figR1} shows the empirical spread of the fractional residuals, defined as $(84^{\rm th} - 16^{\rm th})/2$ percentiles, together with the GRU-predicted 1$\sigma$ uncertainty, as a function of radius for both the training and test samples. The close agreement between the training and test curves, as well as between the empirical fractional residuals and model-predicted errors, indicates stable and well-calibrated uncertainty estimates across radius, thereby demonstrating the robustness of the model. 

Figure~\ref{figA_rec_fine} in Appendix~\ref{mass_reconstruction} shows 40 randomly selected clusters from the unaugmented testing sample, i.e. simulated ICM profiles with the original 40 bin radial resolution spanning [0.05, 2]\,{R}$_{500}$. Later, we also discuss the performance with respect to different input ICM resolutions.

Finally, we test directly whether the bidirectional architecture exploits outer-radius information when predicting inner masses. For every test cluster, we re-evaluate the network on profiles truncated at each evaluation radius $\rm R$, retaining only data interior to that radius, and compare the resulting prediction of $\rm M(<R)$ with the full-input prediction at the same radius. As shown in Figure~\ref{figR1_new}, the truncated-input prediction remains essentially unbiased over most of the radial range: beyond $\rm R \sim  0.5\,R_{500}$ the truncated and full-input curves are indistinguishable, and the median fractional residual stays within a few percent of zero. Only in the central region, $\rm R \lesssim 0.5\,R_{500}$, does the effect of truncation begin to grow, reaching a maximum of $\sim 0.4$ in fractional terms near the centre. This demonstrates that the model's predictions are largely robust to the loss of outer data, and that the residual reliance on larger radii is confined to the innermost region where the gas fraction is most sensitive to feedback physics. For observational applications, radial coverage extending to $\sim \rm 0.5\,R_{500}$ is therefore sufficient to recover the full-profile accuracy at all input radii.

\subsection{Performance with respect to the dynamical states of galaxy clusters}
Since the dynamical state directly influences the properties of the ICM, it is essential to evaluate the GRU model performance across these two distinct populations. We used the GIZMO-SIMBA test sample, for which dynamical state parameters are available (GADGET-X test sample clusters were excluded as we did not estimate their dynamical parameters). The left panel of Fig.~\ref{figR2} shows the median fractional residuals together with the corresponding $84^{\rm th} - 16^{\rm th}$ percentile range for the full unaugmented GIZMO-SIMBA test sample of 1610 galaxy clusters, as well as for the 500 most relaxed and 500 most disturbed clusters within this sample. Here, we also see that the median fractional remains close to 5\% over most of the cluster radial region. These results show that the model maintains high predictive accuracy for both relaxed and disturbed systems. The scatter in M$_{500}$ is 0.021 dex for the full unaugmented  GIZMO-SIMBA test sample, compared to 0.020 dex and 0.022 dex for the 500 most relaxed and 500 most disturbed clusters, respectively, demonstrating the robustness of the model to variations in cluster dynamical state.
\begin{figure}
\centering
		\includegraphics[width=0.4\textwidth]{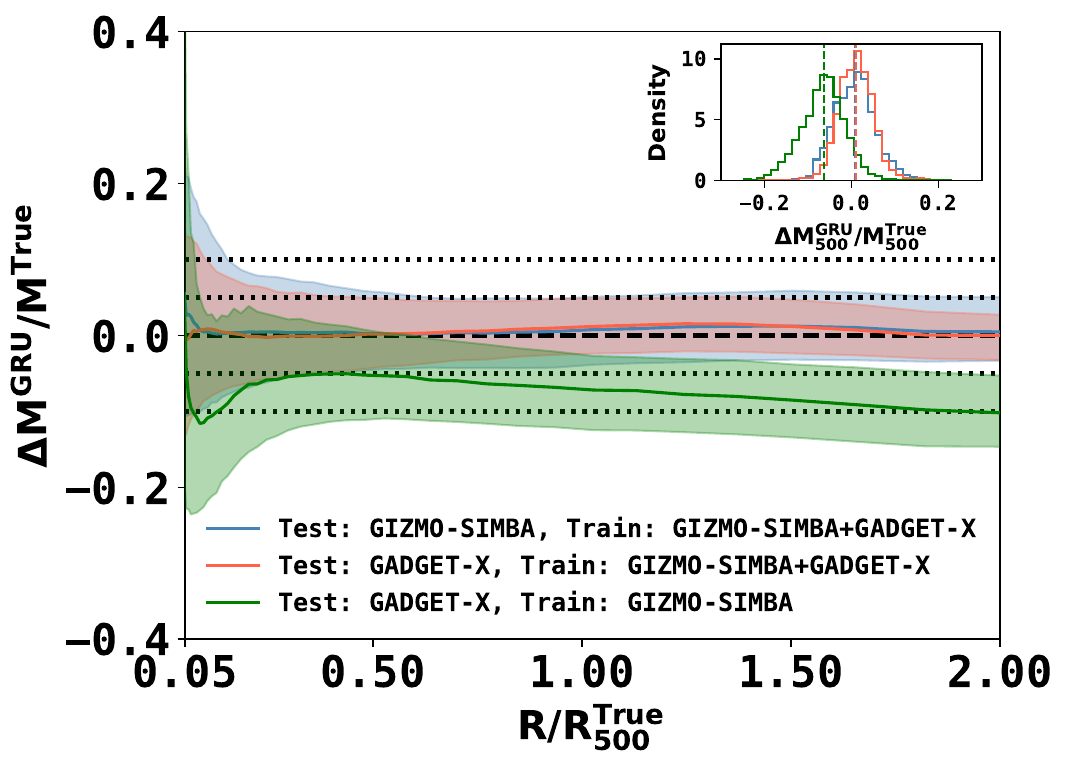}
		\caption{\footnotesize Residuals of the predicted cluster mass profiles relative to the true profiles as a function of normalised radius for unaugmented test samples and models trained on different simulation suites. The solid curves show the median residuals, while the shaded regions indicate the $84^{\rm th}-16^{\rm th}$ percentile interval. Results are shown for three train--test configurations: GIZMO-SIMBA test clusters evaluated with a model trained on the combined GIZMO-SIMBA and GADGET-X training set (blue), GADGET-X clusters evaluated with the same jointly trained model (red), and GADGET-X clusters evaluated with a model trained exclusively on GIZMO-SIMBA (green). The jointly trained model yields nearly unbiased mass profile predictions across both simulation suites, whereas training solely on GIZMO-SIMBA introduces a systematic bias of approximately $5$\%-$10\%$ when applied to GADGET-X clusters. Horizontal dotted black lines represent $\pm$5\% and $\pm$10\% of residuals, and the dashed line represents zero residuals. The inset shows the distribution of fractional residuals in M$_{500}$ for all three cases, with vertical dashed lines indicating the median of each distribution.  }
		\label{figR4bbb}
\end{figure}
\subsection{Reconstruction of mass profiles with an observation-like binning}
In real-world observational cases, mass profiles are reconstructed using non-uniform binning schemes dictated by the resolution and sensitivity of instruments. To align the model's outputs with these practical constraints, we estimated the predicted mass profiles into observation-like bins. These bins are typically logarithmically spaced, considering 14 bins up to R$_{500}$ and 7 bins up to 0.75\,R$_{500}$. The right panel of Fig.~\ref{figR2} presents the median fractional residuals for the observational-like cases as well as for the finely binned unaugmented cases. Despite the reduced resolution and the associated loss of information, the model preserves a comparable level of accuracy i.e. 5\% fractional residuals over most of the cluster range. This result is significant because it demonstrates that the model is not only effective in the idealised simulation domain but also applicable to observational-like scenarios where data are constrained by instrumental and procedural limitations.  We find that the scatter in mass across all clusters is 0.018 at R$_{500}$ and 0.022 at 0.75\,R$_{500}$ for the 14 bins and 7 bins cases, respectively. This adaptability makes the proposed model a practical tool for interpreting observational data, bridging the gap between simulations and real-world observations.
\begin{figure}
\centering
		\includegraphics[width=0.4\textwidth]{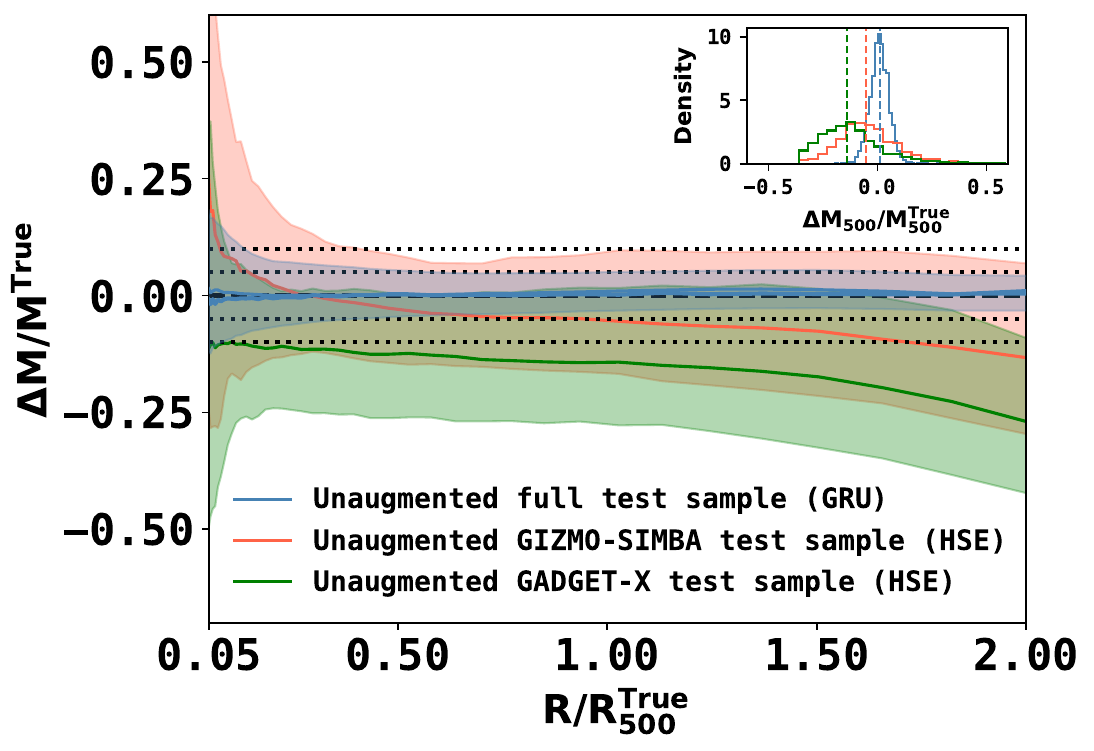}
		\caption{\footnotesize Comparison of GRU-predicted and HSE masses for the test sample as a function of normalised radius. The solid lines represent the median fraction residuals, and the shaded regions indicate the $84^{\rm th}-16^{\rm th}$ percentile interval. The GRU-derived masses are largely unbiased, with a scatter of $\sim$5\%. In contrast, HSE masses exhibit a radially increasing bias, reaching $\sim$23\% at $2\,{\rm R}_{500}$ for the GADGET-X sample, while the GIZMO-SIMBA sample shows a smaller bias of $\sim$13\% at $2\,{\rm R}_{500}$. The inset shows the distribution of fractional residuals in M$_{500}$ for all three cases, with vertical dashed lines indicating the median of each distribution.  Horizontal dotted black lines represent $\pm$5\% and $\pm$10\% of residuals, and the dashed line represents zero residuals. }
		\label{figR3}
\end{figure}
\subsection{Contrasting with a different set of simulation suites}
We also evaluate the robustness of the proposed model by testing its ability to generalise across two independent hydrodynamical simulation suites: GADGET-X and and GIZMO-SIMBA used in this work. When trained jointly on both datasets, the model is able to recover unbiased total mass profiles for clusters from either simulation, indicating that the learned mapping between ICM profiles and total mass is not strongly dependent on the underlying numerical implementation. However, as shown in Fig.~\ref{figR4bbb}, a noticeable degradation in performance is observed when the model is trained exclusively on the GIZMO-SIMBA simulations and evaluated on the GADGET-X  dataset. In this case, the predicted mass profiles exhibit a systematic bias of approximately 5\%-10\%, indicating a mismatch in the learned representation across simulation domains. This suggests that although the ICM–mass relation is physically motivated, residual differences in the gas physics, subgrid modelling, and numerical schemes between the two simulation frameworks introduce domain-dependent features that are captured by the model. These results highlight the importance of incorporating diverse simulation datasets during training to improve robustness.
\begin{figure*}
\sidecaption
		  \includegraphics[width=12.9cm]{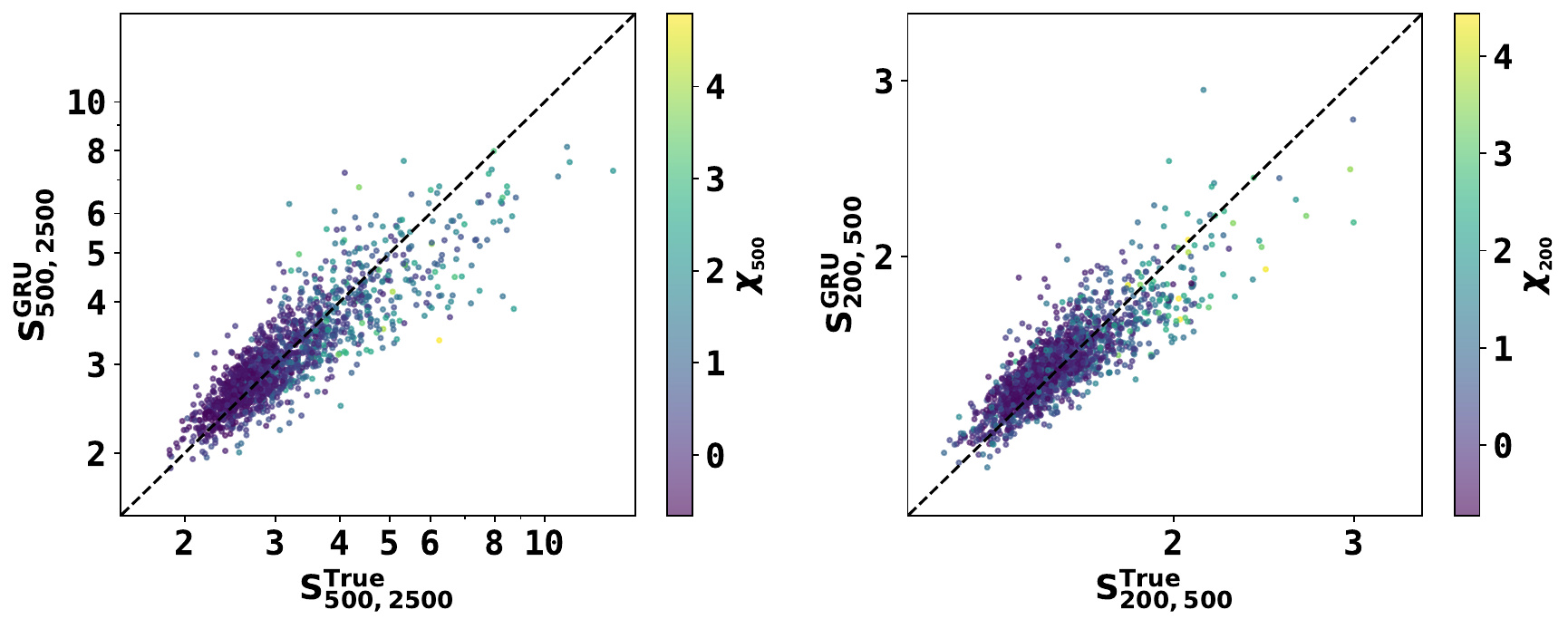}
		\caption{\footnotesize GRU-derived cluster sparsity for the GIZMO-SIMBA test sample. The GRU predicts S$^{\rm GRU}_{200,500}$ and S$^{\rm GRU}_{500,2500}$, with scatters of 0.028 and 0.063 dex, respectively. The larger scatter at smaller radii reflects the more complex inner cluster structure.}
		\label{figR4tt}
\end{figure*}
\begin{figure}
\centering

		\includegraphics[width=0.36\textwidth]{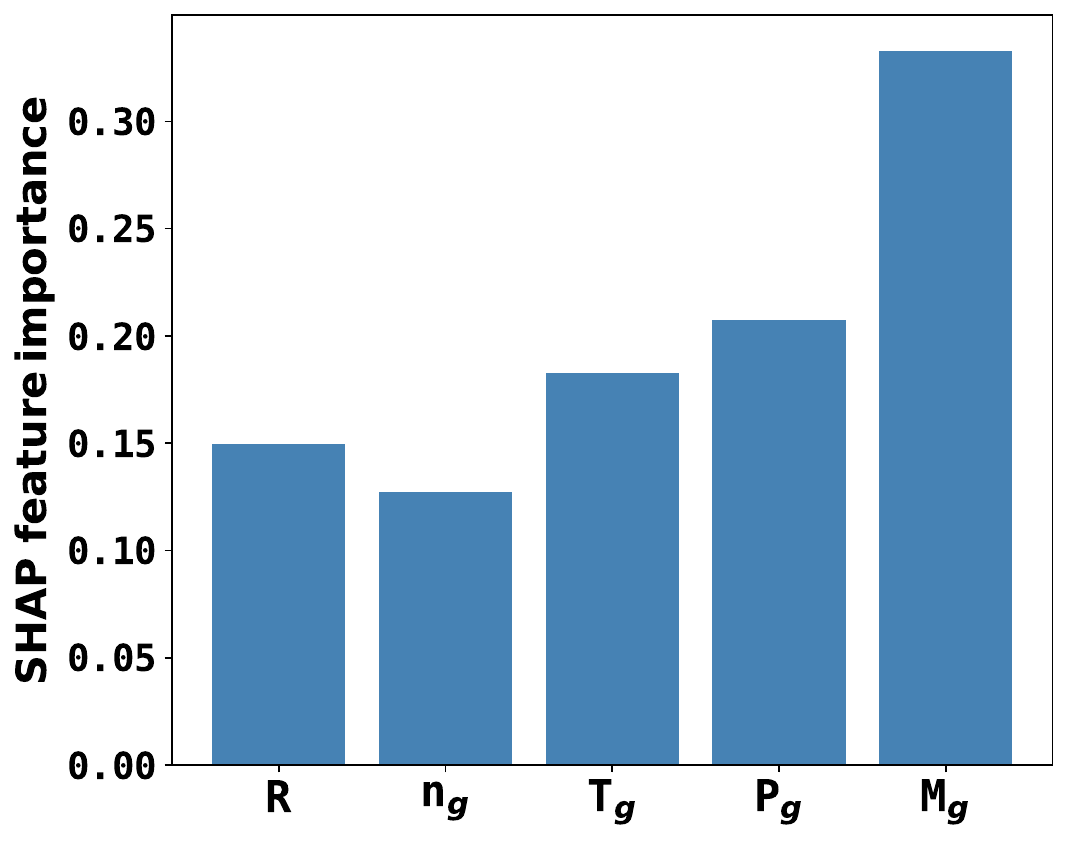}
		\caption{\footnotesize SHAP importance values showing that the enclosed ICM gas mass dominates the prediction of total mass, with pressure, temperature, and density profiles shaping its radial structure. }
		\label{figR4}
\end{figure}
\begin{figure}
\centering

		\includegraphics[width=0.4\textwidth]{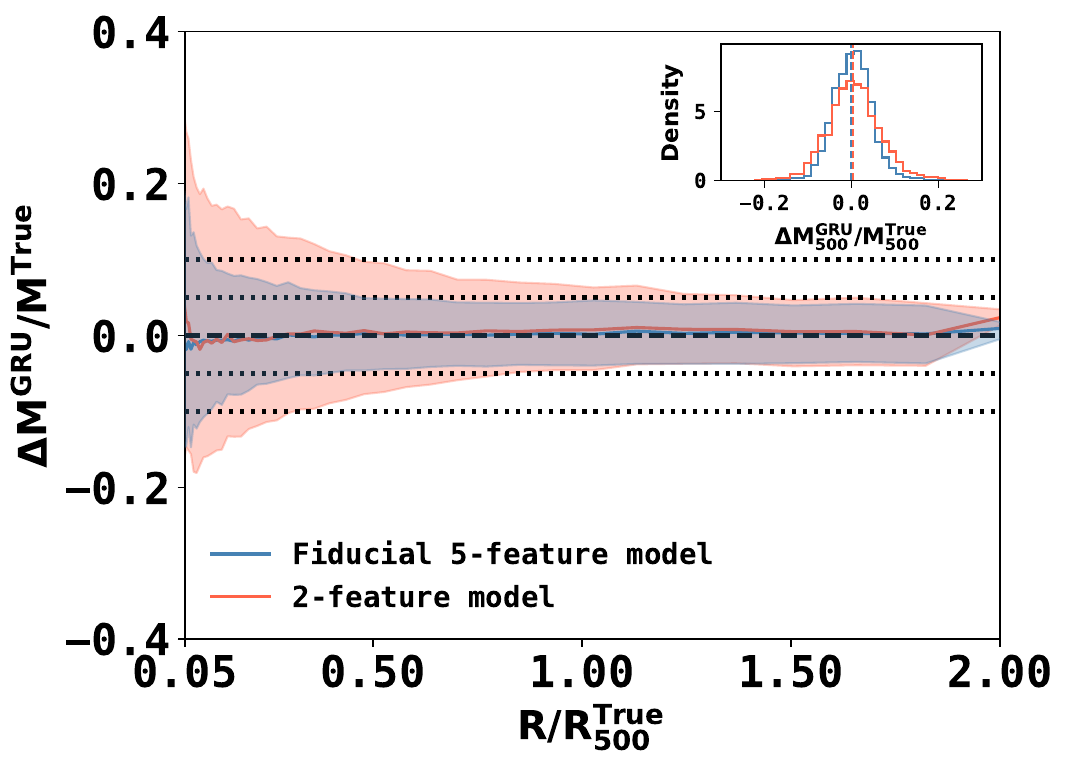}
		\caption{\footnotesize Comparison of the performance of the fiducial five-feature model and the reduced two-feature model (radius and integrated gas mass) is presented in terms of the median fractional residual and the $84^{\rm th}-16^{\rm th}$ percentile range. The inset shows the distribution of fractional residuals in M$_{500}$ for both models, with vertical dashed lines marking the corresponding medians.  Horizontal dotted black lines represent $\pm$5\% and $\pm$10\% of residuals, and the dashed line represents zero residuals.  }
		\label{figR4b}
\end{figure}
\subsection{Contrasting with traditional hydrostatic mass profiles}
 To benchmark the GRU model against traditional HSE methods, we compare the HSE mass profiles of unaugmented test sample to the GRU reconstructed mass profiles. Figure~\ref{figR3} reveals that the GRU model consistently outperforms HSE based estimates, particularly in the outer regions where deviations from equilibrium are prominent. We find that while the GRU-predicted masses are unbiased, with a fractional dispersion of about 5\% over most of the cluster region, the hydrostatic masses are biased, with the bias similar to that found for the full sample discussed in Sec.~\ref{sec:sub2}. In Fig.~\ref{figR4tt}, we also show the GRU-derived sparsity with the true sparsity for the test sample. We find that the scatter in S$^{\rm GRU}_{200,500}$ and S$^{\rm GRU}_{500,2500}$  to be 0.028 and 0.060, respectively. By comparison, in the HSE case, the scatter in the test sample is 0.13 dex and 0.19 dex for S$^{\rm HSE}_{200,500}$ and S$^{\rm HSE}_{500,2500}$, respectively, again highlighting the improved accuracy of the GRU across all radii.

\begin{figure*}
 \centering
\begin{minipage}{0.49\linewidth}
 \includegraphics[width=\textwidth]{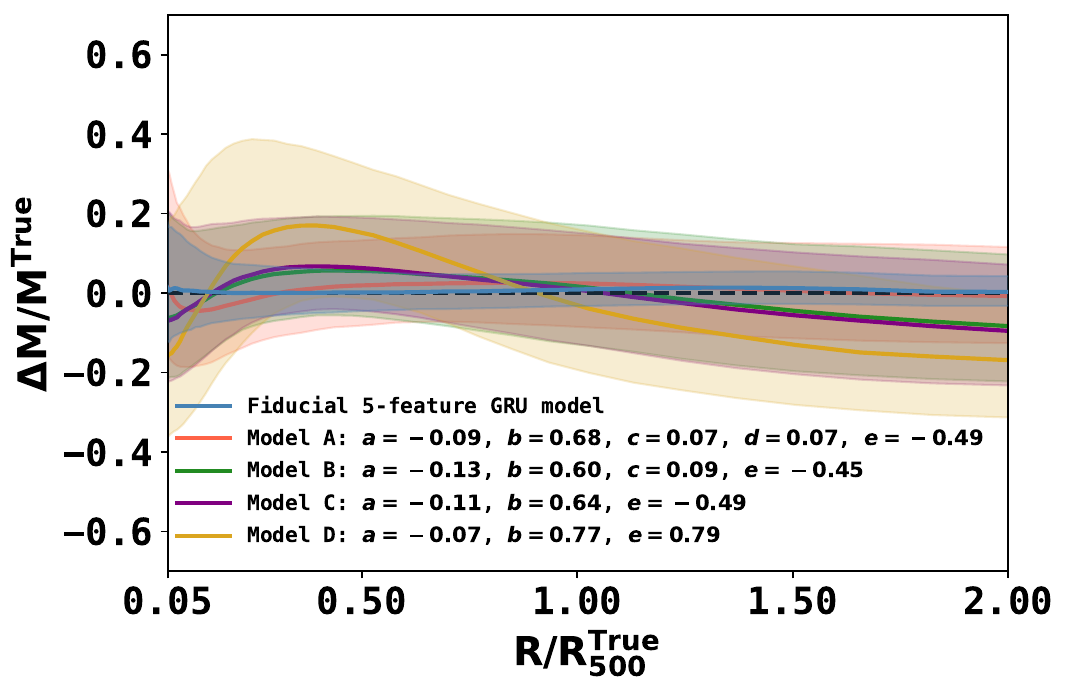} %
\end{minipage}
\begin{minipage}{0.49\linewidth}
\includegraphics[width=\textwidth]{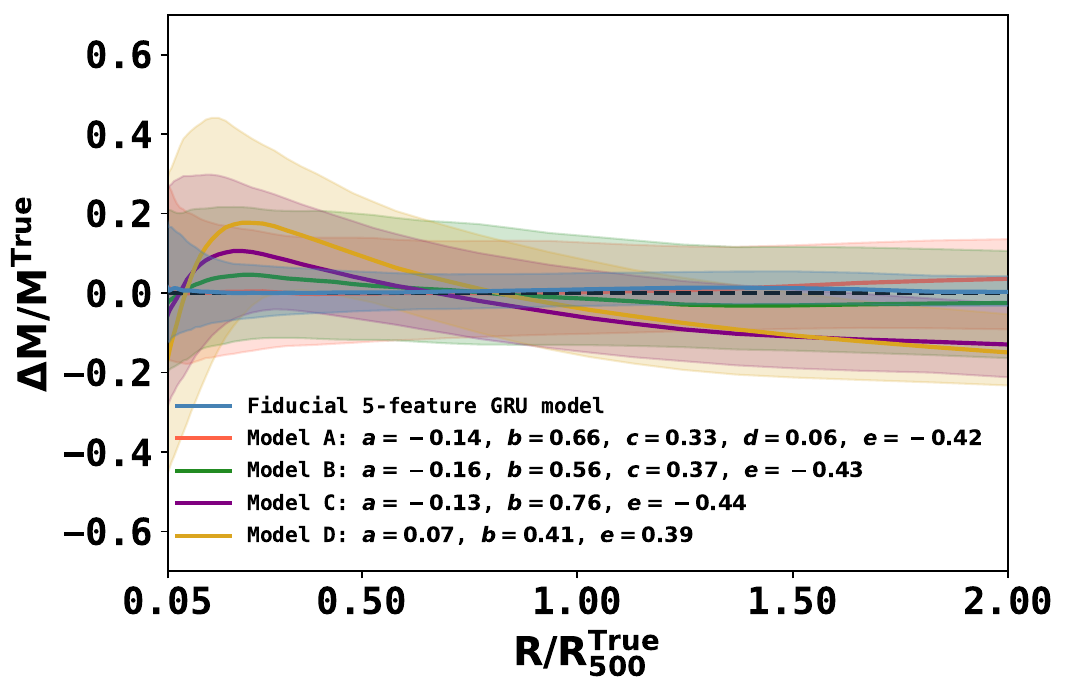}
\end{minipage}
\caption{\footnotesize Median fractional residual as a function of the normalised radius for the fiducial five-feature GRU and four gas-mass--total-mass scaling relations. The three nested relations, listed in order of decreasing complexity, are Model~A [$\log_{10}(\mathrm{M}_{\rm tot}/10^{15}\mathrm{M}_\odot) = a + b\,\log_{10}(\mathrm{M}_{\rm gas}/10^{14}\mathrm{M}_\odot) + c\,\log_{10}(\mathrm{R}/10^3\,{\rm Kpc}) + d\,\log_{10}(\mathrm{R}/10^3\,{\rm Kpc})\,\log_{10}(\mathrm{M}_{\rm gas}/10^{14}\mathrm{M}_\odot) + e\,\log_{10}(1+z)$], Model~B [$\log_{10}(\mathrm{M}_{\rm tot}/10^{15}\mathrm{M}_\odot) = a + b\,\log_{10}(\mathrm{M}_{\rm gas}/10^{14}\mathrm{M}_\odot) + c\,\log_{10}(\mathrm{R}/10^3\,{\rm Kpc}) + e\,\log_{10}(1+z)$], and Model~C [$\log_{10}(\mathrm{M}_{\rm tot}/10^{15}\mathrm{M}_\odot) = a + b\,\log_{10}(\mathrm{M}_{\rm gas}/10^{14}\mathrm{M}_\odot) + e\,\log_{10}(1+z)$], together with the gas-fraction power law, Model~D [$\log_{10}(\mathrm{M}_{\rm gas}/\mathrm{M}_{\rm tot}) = a + b\,\log_{10}(\mathrm{R}/10^3\,{\rm Kpc}) + e\,\log_{10}(1+z)$]. The left panel shows the GIZMO test sample, and the right panel shows the GADGET test sample. Solid lines mark the median of the fractional residual, and shaded regions mark the $84^{\rm th}$--$16^{\rm th}$ percentile range. The best-fit parameters are displayed in the figure legends. The dashed line marks zero residual. The analytic relations remain biased and less accurate than the GRU across the full radial range.}
    \label{fig_new1} 
\end{figure*}
\subsection{Model insights: Attributing predictions to input features} 
Understanding the contribution of individual input features to the model predictions provides insight into the physical processes governing galaxy cluster mass profiles and enhances the transparency of the modelling framework. To quantify feature importance, we employ SHAP \citep[SHapley Additive exPlanations,][]{10.5555/3295222.3295230}, a model-agnostic interpretability method grounded in cooperative game theory. SHAP assigns each input feature a contribution to the model output by computing Shapley values, which represent the average marginal contribution of a feature across all possible feature combinations. This approach provides a consistent and theoretically well-founded measure of feature importance, enabling both global and local interpretability of complex, non-linear models.

Figure \ref{figR4} shows the relative importance of each input feature in predicting the total mass of galaxy clusters. Our SHAP analysis reveals that the gas mass feature is the most important feature for predicting mass profiles, with normalised SHAP importance values of [R, n$_{g}$, T$_{g}$, P$_{g}$, M$_{g}$] = [0.15, 0.13, 0.18, 0.21, 0.33]. The prediction of total integrated mass profiles from ICM observables inherently relies on the cumulative and physically coupled nature of cluster mass distributions. Both the total mass, M(<R), and the ICM gas mass, M$_g$(<R), are enclosed quantities, defined as integrals over radius. The enclosed gas mass traces the baryonic component of the cluster and provides an important constraint on the normalisation and radial evolution of the total mass profile. Because the gas fraction varies only slowly with radius and cluster mass, the gas mass provides a physically motivated anchor for the cumulative mass distribution and limits the range of plausible total mass profiles. Therefore, the prominence of gas mass in the SHAP analysis reflects its fundamental role in constraining the cluster mass distribution rather than the presence of circular or redundant information.

To assess the relative importance of these features, we trained a model using only the radius and gas mass as inputs. The resulting SHAP importance values for this model are [R, M$_{g}$] = [0.34, 0.66], and its performance showed only a slight decline compared to the fiducial five-feature model with SHAP importance values of [R, M$_{g}$] = [0.34, 0.66]. Figure \ref{figR4b} shows the median fractional residuals for both models, highlighting that the two-feature model closely follows the fiducial model beyond the cluster inner region. For the unaugmented test sample, the scatter in M$_{500}$ for the two-feature model is 0.027 dex, compared with 0.019 dex for the fiducial five-feature model. This outcome is consistent with the physical interpretation suggested by the SHAP-based analysis. Pressure, temperature, and density further constrain the total mass distribution by encoding the thermodynamic state of the ICM and its response to the underlying gravitational potential through hydrostatic equilibrium. As the radius determines the spatial scale over which physical properties such as density and temperature are measured, its influence is inherently reflected in the corresponding SHAP values.

The reduced GRU provides a direct test of whether the model's accuracy is attributable to the additional thermodynamic channels or to the sequence architecture itself. To isolate the role of the network, we compare the GRU models against a hierarchy of four gas-mass--total-mass scaling relations, all fitted to the same training set, evaluated on the same test set, and all including $\log_{10}(1+z)$ as a predictor. Three of these form a nested sequence in which each relation is a simplification of the previous one: Model~A is the most general form and includes a gas-mass--radius interaction term, $\log_{10}({\rm M}_{\rm tot}/10^{15}\,{\rm M}_\odot) = a + b\,\log_{10}({\rm M}_{\rm gas}/10^{14}\,{\rm M}_\odot) + c\,\log_{10}({\rm R}/10^{3}\,{\rm Kpc}) + d\,\log_{10}({\rm R}/10^{3}\,{\rm Kpc})\,\log_{10}({\rm M}_{\rm gas}/10^{14}\,{\rm M}_\odot) + e\,\log_{10}(1+z)$; Model~B drops the interaction term and retains a radial power law, $\log_{10}({\rm M}_{\rm tot}/10^{15}\,{\rm M}_\odot) = a + b\,\log_{10}({\rm M}_{\rm gas}/10^{14}\,{\rm M}_\odot) + c\,\log_{10}({\rm R}/10^{3}\,{\rm Kpc}) + e\,\log_{10}(1+z)$; and Model~C drops the radial dependence entirely, $\log_{10}({\rm M}_{\rm tot}/10^{15}\,{\rm M}_\odot) = a + b\,\log_{10}({\rm M}_{\rm gas}/10^{14}\,{\rm M}_\odot) + e\,\log_{10}(1+z)$. The fourth, Model~D, is the gas-fraction power law $\log_{10}({\rm M}_{\rm gas}/{\rm M}_{\rm tot}) = a + b\,\log_{10}({\rm R}/10^{3}\,{\rm Kpc}) + e\,\log_{10}(1+z)$.

Figure~\ref{fig_new1} compares the median fractional residual for the fiducial five-feature GRU together with all four scaling relations as a function of radius, with the GIZMO-SIMBA test sample and the GADGET-X test sample, which are known to produce different baryon profiles, especially in the cluster inner regions \citep{2023MNRAS.523.1228L}. For the GIZMO-SIMBA test sample, Model~A achieves a total scatter over clusters and over all radii beyond 0.1\, R$_{500}$ of $0.060$ dex ($14.7\%$), Model~B yields $0.066$ dex ($16.5\%$), Model~C $0.066$ dex ($16.4\%$), and Model~D is substantially broader at $0.099$ dex ($25.7\%$). On the GADGET-X test sample, the same sequence gives $0.072$ dex ($18.3\%$), $0.077$ dex ($19.4\%$), $0.091$ dex ($23.4\%$), and $0.131$ dex ($35.3\%$), respectively.  A joint fit on the combined GIZMO-SIMBA$+$GADGET-X training set performs systematically worse than either simulation-specific fit, with scatters over the full test sample of $0.096$~dex ($24.7\%$) for Model~A, $0.097$~ ex ($25.0\%$) for Model~B, $0.125$ dex ($33.6\%$) for Model~C, and $0.197$ dex ($57.6\%$) for Model~D, confirming that a single set of coefficients cannot capture the simulation-dependent differences in the gas-mass--total-mass relation. The improvement from Model~C through Model~A is modest compared with the gap to the network. For both the fiducial five-feature GRU and the reduced two-feature GRU, the total scatter over all radii is dominated by the lower radii because of the logarithmic binning: the five-feature model yields $\sim 0.030~\mathrm{dex}$ ($\sim 8\%$), and the two-feature model yields $\sim 0.050~\mathrm{dex}$ ($\sim 12\%$). Unlike the power-law scaling relations, both GRU models achieve fractional scatter below $10\%$ beyond the cluster central region, decreasing to $\sim\!5\%$ at $\mathrm{R}_{500}$. The gain therefore does not arise solely from the additional thermodynamic channels: the sequence-to-sequence architecture itself learns a non-linear, radius-dependent mapping from $(\rm R, M_{\rm gas})$ to $\rm M_{\rm tot}(<R)$ that a fixed power-law parameterisation cannot represent. This isolates the contribution of the deep learning methodology from that of the input features, and shows that the improvement is intrinsic to the learned representation rather than reducible to any single physical channel.
\section{Application to observed data}
We utilised two complementary samples of galaxy clusters to investigate their mass distribution and thermodynamic properties using a GRU-based modelling approach.  The REXCESS (Representative XMM-Newton Cluster Structure Survey) sample comprises of 31 clusters drawn from the REFLEX catalogue, selected to provide a statistically representative coverage of cluster masses and morphologies at low redshift. X-ray observations from XMM-Newton enabled detailed mapping of the ICM, including radial profiles of density, temperature, and entropy \citep{Bohringer2007,Pratt2010,2010A&A...517A..92A} up to R$_{500}$. These profiles characterised both the structural and thermodynamic properties of clusters and underpin scaling relations that link X-ray observables, such as luminosity and temperature, to cluster hydrostatic mass \citep{2008A&A...487..431C,2009A&A...498..361P,2010A&A...517A..92A}. Prior analyses of REXCESS have also examined the effects of non-gravitational feedback processes \citep{Iqbal2018}.  The X-COP (XMM-Newton Cluster Outskirts Project) sample, in contrast, consists of 12  X-ray and SZ-selected clusters designed to probe cluster properties out to their outskirts by combining X-ray data with SZ observations \citep{2017AN....338..293E,2019A&A...621A..41G,2019A&A...621A..40E}. These measurements provide important insights into hydrostatic mass, cluster formation and evolution \citep{2019A&A...621A..41G,2019A&A...621A..39E,2022A&A...662A.123E}.

For both samples, GRU mass profiles were derived by propagating observational uncertainties in the gas density and temperature profiles through an outer Monte Carlo (MC) loop. For each cluster, we generate $N_{\mathrm{obs}} = 100$ realisations of the observed profiles by sampling from log-normal distributions \citep{2007ApJ...659..257K}. Derived thermodynamic profiles (pressure, integrated gas mass) and gas mass profiles are recomputed for each realisation and passed through the neural network ensemble. For each realisation, model uncertainty is then propagated using another MC loop: mass profiles and corresponding uncertainties are predicted using an ensemble of 50 GRU-based models, collapsed into a single predictive distribution in logarithmic mass space via weighted averaging. From this distribution, $N_{\mathrm{model}} = 100$ realisations of mass profiles are drawn. As a result, each cluster has 100 $\times$ 100 = 10,000 mass profile realisations. Final profiles are summarised using the median and the 16th–84th percentiles of these aggregated samples at each radius, yielding credible intervals that consistently incorporate uncertainties.

\begin{figure*}
	\sidecaption
	\begin{minipage}[b]{0.35\textwidth}
		\includegraphics[width=\textwidth]{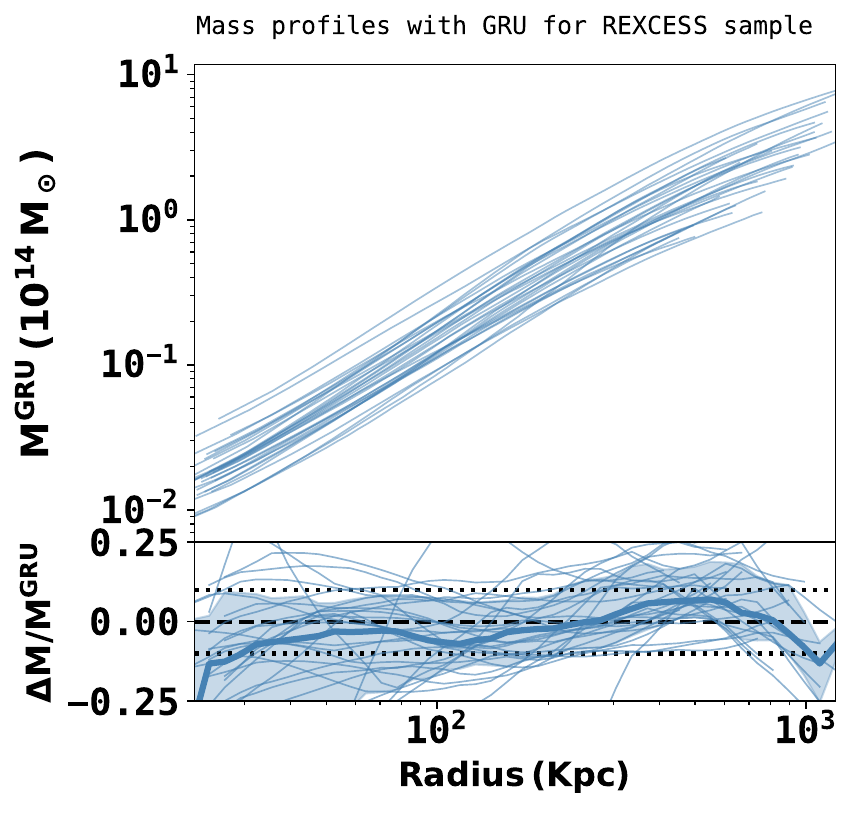}
	\end{minipage}
	\hfill
	\begin{minipage}[b]{0.35\textwidth}
		\includegraphics[width=\textwidth]{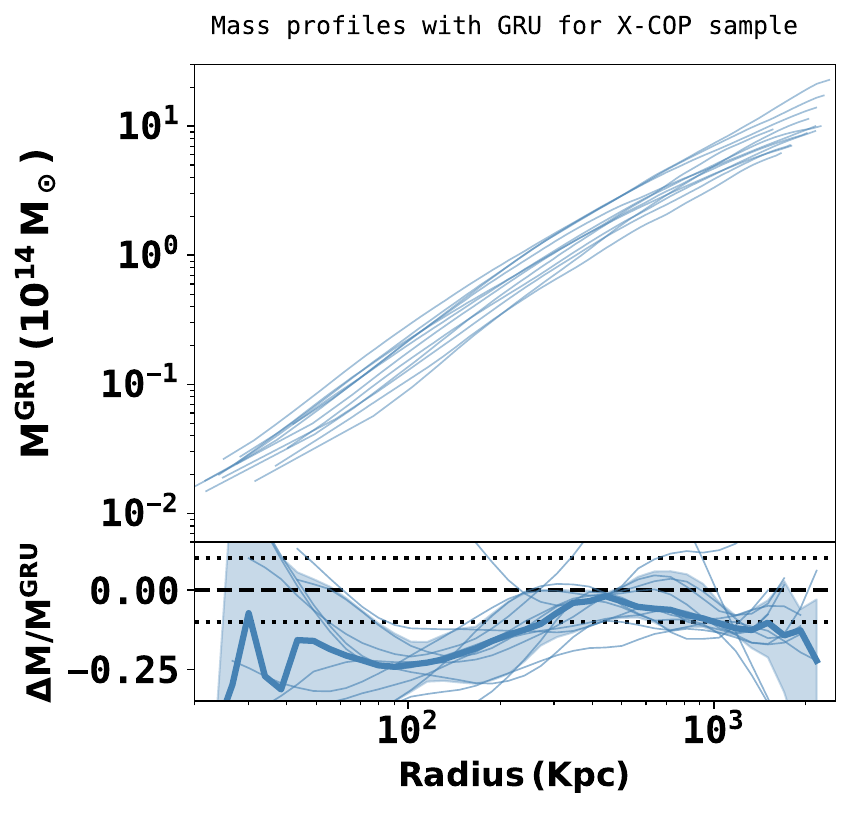}
	\end{minipage}
	\caption{\footnotesize
		Cumulative mass profile reconstruction for the REXCESS sample (left panel) and the X-COP sample (right panel) using the GRU model. For clarity, the error envelopes of the individual profiles are omitted here and shown instead in Figs.~\ref{figA_REXCESS} and \ref{figA_XCOP} of Appendices~\ref{appREXCESS}. In both panels, the lower subplots show the fractional residuals of the hydrostatic masses relative to the GRU masses, \(({\rm M}^{\rm HSE}-{\rm M}^{\rm GRU})/{\rm M}^{\rm GRU}\).
		The thick blue line and shaded region denote the median fractional residual and the \(84^{\rm th}\)--\(16^{\rm th}\) percentile range, respectively, for the best-fitting model of the full sample. Horizontal dotted black lines represent \(\pm10\%\) residuals, while the dashed horizontal line marks zero residual.
	}
	\label{fig10}
\end{figure*}

\begin{figure*}
	\sidecaption
	\includegraphics[width=0.70\textwidth]{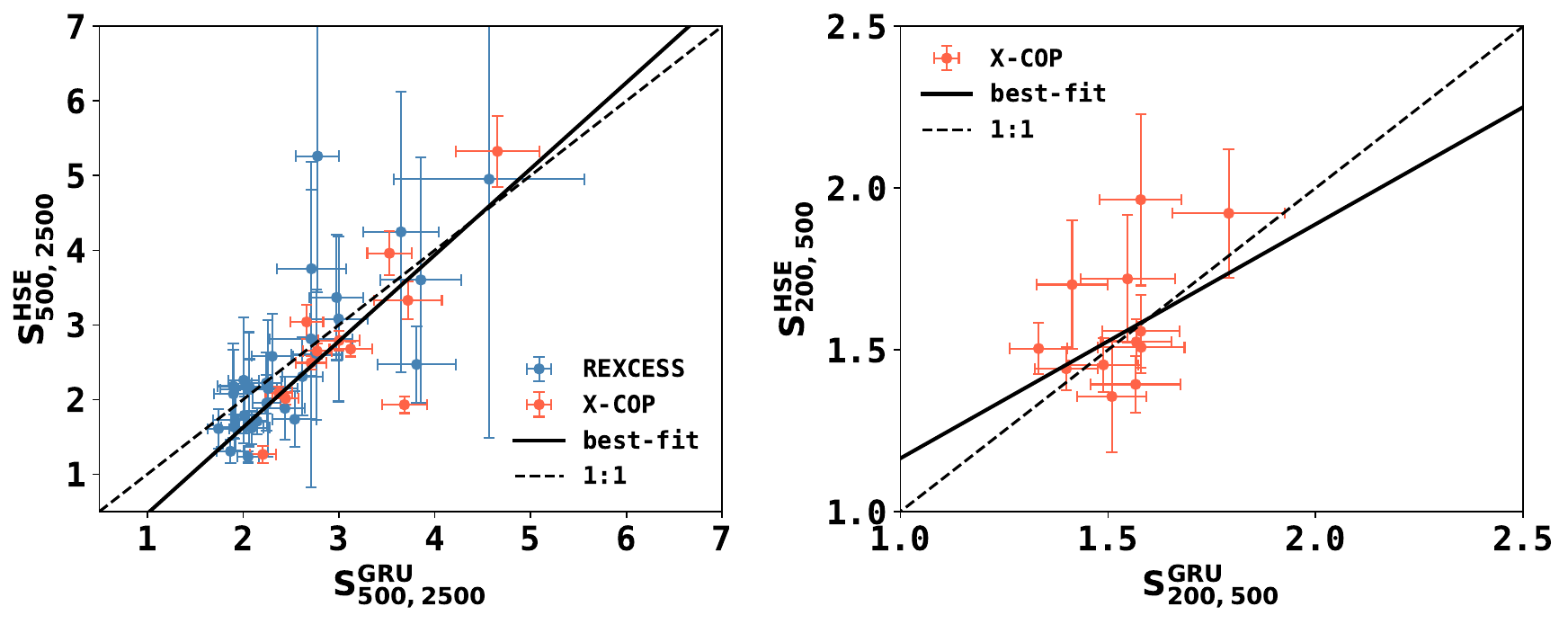}
	\caption{\footnotesize Comparison between HSE- and GRU-derived cluster sparsity estimates. The left panel shows S\(_{500,2500}\) for the combined REXCESS (blue) and X-COP (red) samples, while the right panel presents S\(_{200,500}\) for the X-COP sample alone. Error bars denote the \(1\sigma\) uncertainties. The dashed black line indicates the one-to-one relation, and the solid lines represent the best-fitting linear relations.}
	\label{figR4ttt}
\end{figure*}

Figure~\ref{fig10} presents the cumulative mass profiles derived from the GRU model for the REXCESS and X-COP cluster samples, together with the fractional residuals relative to the HSE mass estimates. The reconstructed mass profiles show the expected monotonic increase with radius. For the REXCESS clusters, the median residual remains close to zero over most radii, indicating good agreement between the GRU and HSE mass estimates. In contrast, the X-COP sample, which comprises systematically more massive clusters, exhibits predominantly negative residuals, corresponding to GRU-predicted masses that are higher than the HSE estimates by up to approximately 10\%.

The higher GRU masses for the more massive X-COP clusters are consistent with the expected direction and magnitude of hydrostatic mass bias reported in previous simulation and observational studies. However, this offset should not be interpreted as a direct measurement of hydrostatic bias. In the absence of an independent mass calibration, such as weak-lensing measurements, the difference between the GRU and HSE masses may also arise from systematic differences between the simulations \citep{2023MNRAS.523.1228L} and observations and differences in analysis (X-ray-only analysis or joint X-ray+SZ analysis). Moreover, in the present analysis, the simulated profiles are adapted to the radial sampling of the observations primarily by reducing the number of radial bins, but this procedure does not account for effects such as projection, spectroscopic temperature biases, and gas clumping correction. The potential impact of these differences is also illustrated by the comparison of the two simulation suites used in this study, which
lead to a mass difference of 5–10\% as shown in the previous section. Future work incorporating more realistic mock two-dimensional X-ray observations and independent weak gravitational lensing mass calibration will be important for disentangling hydrostatic bias from simulation--observation differences.

Figure \ref{figR4ttt} compares the cluster sparsities derived from the GRU and HSE methods. For S\(_{500,2500}\), the GRU- and HSE-derived sparsities are positively correlated, although the observational data exhibit substantial scatter. Nevertheless, the best-fitting relation obtained from the combined REXCESS and X-COP samples reveals a mild systematic trend at low sparsities (S\(_{500,2500}\lesssim5\)), consistent with that found in the simulations, where the HSE-derived sparsities are systematically lower than the corresponding true estimates. For the sparsity S\(_{200,500}\), computed using only 12 X-COP clusters, the GRU and HSE estimates also exhibit a positive correlation. However, unlike the simulations, the best-fitting relation shows the opposite systematic behaviour, with the HSE-derived sparsities generally exceeding those predicted by the GRU. Owing to the limited sample size and relatively large intrinsic scatter, this should therefore be interpreted with caution.

Tables~\ref{appx1} and \ref{appx2} in Appendix~\ref{appREXCESS} provide the detailed mass (M$_{500}$ and M$_{200}$) and sparsity estimates for the REXCESS and X-COP cluster samples, respectively. These tables list the SZ-based masses, the graph neural network (GNN) predictions from \citet{Iqbal2025}, and the GRU-derived quantities obtained in this work, together with their associated uncertainties. Finally, Fig.~\ref{figA_XCOPBB} in Appendix \ref{appREXCESS} compares the M$_{500}$ estimates from the GNN model (M$_{500}^{\mathrm{GNN}}$); \citealt{Iqbal2025}) with those from the GRU model (M$_{500}^{\mathrm{GRU}}$) developed in this work for the REXCESS and X-COP samples. The GNN model, introduced by \citet{Iqbal2025}, represents the ICM radial profiles as graphs and learns the relationship between their thermodynamic structure to predict M$_{500}$. The two models show good overall agreement, with most clusters lying close to the one-to-one relation. Notably, the GNN model was trained using only three GADGET-X simulation snapshots, whereas the GRU model was trained on a much larger and more diverse set of simulation snapshots and clusters.
\section{Discussion and conclusions}
In this work, we have demonstrated that ICM thermal profiles contain sufficient information to recover the total true mass distribution of galaxy clusters using a GRU-based approach. By exploiting the full radial structure of the ICM, our method provides a robust alternative to conventional HSE-based estimates, which are sensitive to noisy derivatives and assumptions of hydrostatic equilibrium (bias). The primary findings of this study are summarised as follows:
\begin{itemize}
    \item The GRU model trained on idealised three-dimensional radial profiles exhibits consistent performance across the training and testing datasets, demonstrating good generalisation within the simulation domain. For clusters with total masses in the range $5 \times 10^{13}\, \mathrm{M}_\odot\lesssim {\rm M}_{\mathrm{500}} \lesssim 2.6 \times 10^{15}\, \mathrm{M}_\odot$, the model achieves a median residual fractional dispersion of approximately 5\% over most of the cluster region. Within central regions, a higher residual fractional dispersion of 17\% is found and is plausibly attributable to feedback processes that weaken the correspondence between the ICM and the dark matter density profiles. The GRU model consistently outperforms hydrostatic methods, which exhibit systematic biases.
    \item Across both relaxed and disturbed clusters in the simulated test set, the GRU model delivers nearly identical accuracy and error behaviour. This consistency indicates that the model successfully learns the dominant structural signatures of the ICM, regardless of a cluster’s dynamical state.
    \item Tests with varied resolutions, designed to mimic observational X-ray binning schemes, show negligible degradation in GRU performance, even when data are limited to radii of $0.75\, {\rm R}_{500}$ or ${\rm R}_{500}$ compared to high-resolution simulation data. This demonstrates the robustness of the GRU under different observational constraints.
   \item Feature importance analysis using SHAP identified the ICM gas mass as the most influential predictor of total cumulative mass profiles. This is likely because the gas mass is an integrated quantity encoding cumulative information about the cluster over a range of radii, thus making it intrinsically more informative than local thermodynamic properties. While temperature and pressure also contribute, their impact is secondary.
   \item Joint training on GIZMO-SIMBA and GADGET-X yielded unbiased mass predictions across both simulations, whereas training on a single simulation led to a 5\%-10\% cross-domain bias, highlighting the importance of diverse training. 
    \item For observed clusters, the GRU predictions agree closely with HSE masses for REXCESS, while they are systematically higher than the HSE estimates by up to $\sim$10\% for the more massive X-COP sample. This offset is consistent with the expected direction and magnitude of hydrostatic mass bias but cannot be uniquely attributed to hydrostatic bias in the absence of an independent mass calibration. Differences between simulated and observed profiles as well as differences in the analysis procedures applied to the REXCESS and X-COP samples may also contribute to the observed offset.
\end{itemize}
Apart from these findings, we also explored long short-term memory-based and transformer-based models, and we found a broadly comparable predictive performance. This suggests that the information contained in the ICM profiles is effectively captured by a range of sequential architectures.

Looking ahead, this work can be extended in several directions. Incorporating additional cluster observables, such as SZ effect measurements, shear profiles, or galaxy kinematics, could further constrain mass estimates. Expanding the training dataset to include a wider range of simulation suites is expected to improve model generalisability.  Additionally, alternative mappings using ICM properties, such as differential mass profiles, total density distributions, or non-thermal pressure, could broaden the model’s applicability and provide deeper insight into the connection between thermodynamic structure and the underlying gravitational potential. In this context, the proposed framework can be directly applied to clusters in the CHEX-MATE project \citep[Cluster HEritage project with XMM-{\it Newton}: Mass Assembly and Thermodynamics at the Endpoint of structure formation;][]{2021A&A...650A.104C} by leveraging their observed ICM radial profiles to infer cluster masses and compare physical properties against simulated counterparts. As demonstrated in this work, the GRU model can distinguish between the GIZMO and GADGET samples and reconstruct individual cluster profiles in a joint training setup. This suggests a more classical application in which the learned representation can be used to infer which simulation suite a given ICM most closely resembles, providing a potential framework for testing different models against observations. Future work should also include more realistic mock X-ray observations and independent mass calibrations, such as weak-lensing measurements, to assess the model under observational conditions and to disentangle simulation--observation differences from hydrostatic mass bias. Therefore, combining increasingly diverse simulations with multi-wavelength observations could provide a pathway towards identifying the physical models that best reproduce the observed properties of the ICM. More complex architectures, such as hybrid models combining GNNs with GRUs, could further improve performance by capturing both local structural relationships and global radial dependencies. In such a framework, the GNN model can help encode multi-scale spatial relationships within clusters, while the GRU is used to model the radial and dynamical dependencies, integrating fine-grained local features with cluster-scale trends.
\section*{Data availability}
The GRU model developed in this study is available upon request from the corresponding author. The simulations used for training the model are derived from \thethree   \,(\url{https://www.nottingham.ac.uk/astronomy/The300/index.php}). We use data from the REXCESS project, including detailed X-ray observational profiles presented in \citet{Bohringer2007,2008A&A...487..431C, Pratt2010, 2010A&A...517A..92A}, which were kindly provided by the project team. Additionally, the X-COP data and derived products are publicly accessible online (\url{https://dominiqueeckert.wixsite.com/xcop/about-x-cop}).
\begin{acknowledgements}
We thank the reviewer for their careful reading of the manuscript and for their valuable comments and suggestions, which have helped improve the quality and clarity of our work.
The simulations were performed at the MareNostrum Supercomputer of the BSC-CNS through The Red Espa˜nola de Supercomputaci´on grants (AECT-2022-3- 0027, AECT-2023-1-0013), and at the DIaL – DiRAC machines at the University of Leicester through the RAC15 grant: Seedcorn/ACTP317. We gratefully acknowledge the GPU-enabled high-performance computing resources at the University of Lille for supporting the computational work in this study.  ER acknowledge financial support from NASA grants 80NSSC25K0006 and 80NSSC25K8009. DdA and WC thank Agencia Estatal de Investigación, Ministerio de Ciencia, Innovación y Universidades, Gobierno de España, for financial support from grant REF PID2024-156100NBC21. GWP acknowledges long-term financial support from CNES, the French space agency.
 \end{acknowledgements}

\bibliographystyle{aa}
\bibliography{example} 
\appendix
\section{Simulated ICM profiles in \thethree}
\label{app00}
\begin{figure*}[!htbp]
	\sidecaption
	\includegraphics[width=0.70\textwidth]{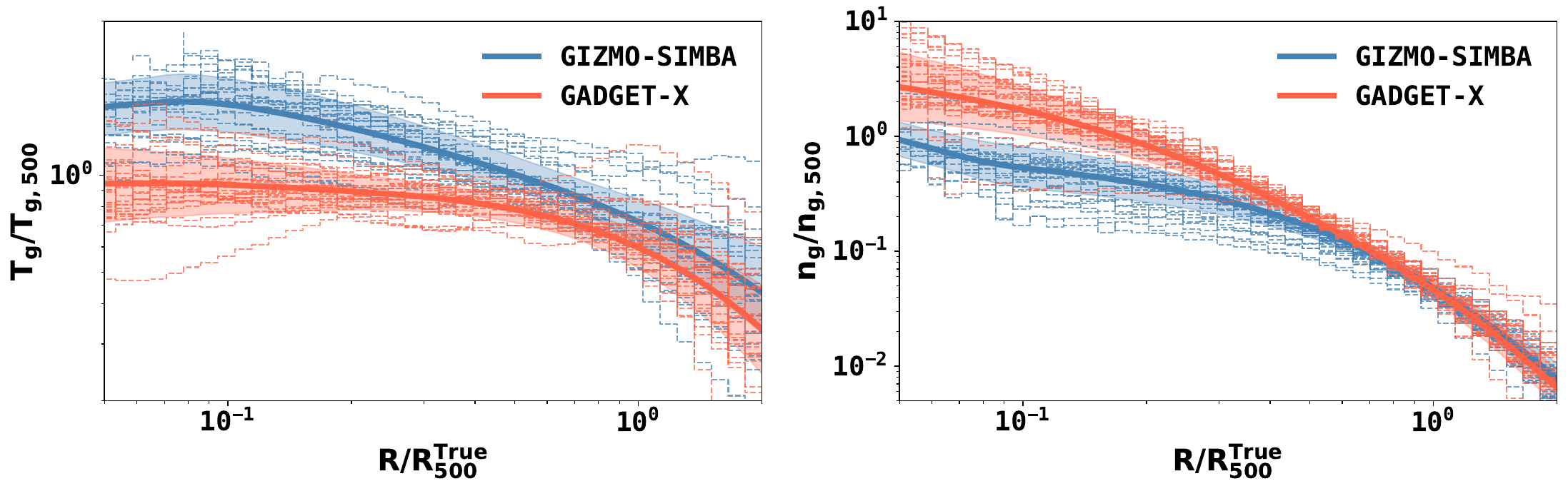}
	\caption{\footnotesize Comparison of  ICM temperature and density profiles in the GIZMO-SIMBA and GADGET-X simulations. The thin dashed lines represent the scaled ICM temperature (left panel) and density (right panel) profiles of 25 randomly selected clusters from the GIZMO-SIMBA (blue) and GADGET-X (red) simulations. The thick solid lines show the corresponding mean profiles of the full sample, while the shaded regions indicate the $84^{\rm th}-16^{\rm th}$ percentiles.
	}
	\label{figA1}
\end{figure*}

Figure~\ref{figA1} presents the ICM temperature and density profiles derived using the two hydrodynamic solvers used in \thethree: GADGET-X (an improved SPH scheme with thermal AGN feedback) and GIZMO-SIMBA (Meshless Finite Mass, MFM, with kinetic AGN outflows). The profiles are scaled by the characteristic temperature T$_{g,500} = 8.85 \times \left(\frac{{\rm M}^{\rm True}_{500}}{10^{15} {\rm M}_\odot}\right)^{2/3} E(z)^{2/3} \left(\mu/0.6\right)$, and characteristic density n$_{g,500} = 500 \times \rho_{\mathrm{crit}}(z)/\mu m_p$, where $\mu=0.6$ and $m_{p}$ is the mass of proton. The gravitational softening lengths are $6.5\,h^{-1}\,\mathrm{kpc}$ for GADGET-X and $5\,h^{-1}\,\mathrm{kpc}$ for GIZMO-SIMBA. Both simulations adopt identical initial conditions and assume a Planck 2016 cosmology \citep{cui2018}. GIZMO‑SIMBA's stronger kinetic feedback systematically reduces central gas density while producing a higher gas temperature compared to GADGET‑X. GADGET‑X exhibits a power‑law entropy core slope consistent with low‑redshift X‑ray observations, whereas GIZMO‑SIMBA produces higher, flatter entropy profiles. However, both simulations agree with observations in outer regions (\({\rm R}\gtrsim0.3\,{\rm R}_{500}\)). For a full quantitative comparison, see \citet{2023MNRAS.523.1228L}.

\section{Pressure profiles of low and high hydrostatic mass bias galaxy clusters}\label{app_pressure_bias}
\begin{figure*}[!htbp]
	\sidecaption
	\includegraphics[width=0.70\textwidth]{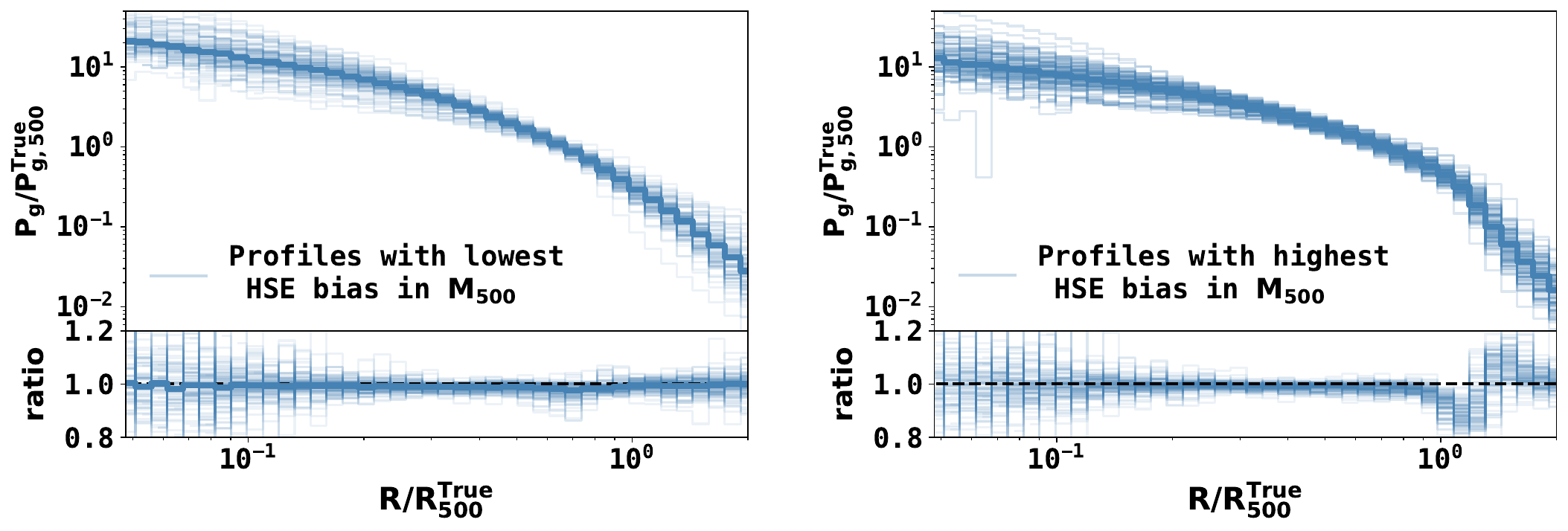}
	\caption{\footnotesize  Pressure profiles of clusters with the 100 lowest (left panel) and 100 highest (right panel) absolute hydrostatic mass bias in M\(_{500}\). The lower panels present the ratio of reconstructed to true pressure profiles. 
	}
	\label{figA2}
\end{figure*}

Figure~\ref{figA2} compares the pressure profiles (scaled by ${\rm P}_{500}^{\mathrm{}} = 1.65 \times 10^{-3} \, E(z)^{8/3} \left( \frac{{\rm M}_{500}^{\mathrm{True}}}{3 \times 10^{14} \, {\rm M}_\odot} \right)^{2/3} \, \mathrm{keV \, cm^{-3}}$) of 100 clusters with the lowest and highest absolute hydrostatic mass bias in our sample. In each case, we show both the true pressure profiles and the corresponding reconstructed profiles obtained using our local fitting scheme. Overall, for both cases, the reconstructed pressure profiles closely follow the true profiles across the radial range, with typical deviations remaining within $\sim 10\%$. 

\section{Summary of GRU model architecture and training Hyper-parameters}
\label{app000}
This section provides additional details on the GRU architecture and the hyper-parameters used during training. Table~\ref{tab:gru_architecture} presents the complete network architecture, including the input and output dimensions of each layer and a brief description of its function within the model. Table~\ref{tab:hyperparameters} summarises the optimised training hyper-parameters, including the learning rate, regularisation parameters, and learning-rate scheduling strategy.
    \begin{table}[!hb]
    \scriptsize
    \centering
    \begin{threeparttable}
    	    \caption{\footnotesize Hyper-parameters for  GRU model.}
    \begin{tabular}{ccl}
      \toprule
\toprule
        \textbf{Hyper-parameter} &\textbf{Range/Values considered}& \textbf{Best-fit}  \\
        \midrule
        Learning rate & $10^{-5}-10^{-2}$& \(0.0007\)  \\ \hline
        Weight decay & $10^{-3}-10^{-1}$&  \(0.001\)  \\ \hline
        Learning rate factor& [0.1, 0.3, 0.5, 0.7] & \(0.5\)  \\ \hline
        hidden size &[80, 100, 128]& \(120\)  \\ \hline
        Dropout probability & [0.1,0.2,0.3, 0.4] & \(0.1\)\\ \hline
        Batch size & [32, 64, 128]& \(64\)  \\ \hline
        Optimiser & {\tt Adam} & Fixed \\ \hline
        Scheduler & {\tt ReduceLROnPlateau} & Fixed \\ \hline
        GRU layers& 1-4 & \(2\)  \\ \hline
        Bidirectional & [True, False] &  True \\ \hline
        Epochs & \(400\) & Fixed \\
       \bottomrule
    \end{tabular}
\tablefoot{ Optimised hyper-parameters for our fiducial five-feature GRU model.}
    \label{tab:hyperparameters}
    \end{threeparttable}
\end{table}

\begin{table*}[!htbp]
\scriptsize
\centering
\begin{threeparttable}
\caption{\footnotesize Summary of the fiducial GRU architecture used to predict cumulative total mass profiles of clusters from ICM radial profiles.}
\begin{tabular}{c c c c p{9cm}}
\toprule
\toprule
\textbf{Layer} & \textbf{Type} & \textbf{Input size} & \textbf{Output size} & \textbf{Description} \\
& & \textbf{Bins $\times$ Features} & \textbf{Bins $\times$ Features} & \\
\midrule
1 & Input Sequence &
$40 \times 5$ &
$40 \times 5$ &
Each bin contains five features: ICM temperature, density, pressure, enclosed gas mass, and radius. Profiles containing fewer than 40 bins are padded to a fixed length.\\ 
\hline
2 & Masking / Packing &
$40 \times 5$ &
$N \times 5$ &
Removes padded bins using packed-sequence processing with masking, ensuring only physical bins contribute to the GRU computations, where $N$ denotes the true number of bins. \\
\hline
3 & Bidirectional GRU &
$N \times 5$ &
$N \times 240$ &
Two-layer bidirectional GRU with hidden dimension 120 in each direction learns dependencies between bins and produces a 240-dimensional latent representation for each bin. \\
\hline
4 & Unpacking &
$N \times 240$ &
$40 \times 240$ &
Restores the packed sequences to the original padded length, enabling subsequent operations to be applied to tensors with a uniform shape across the batch. \\
\hline
5 & Redshift Projection &
$1 \times 1$ &
$1 \times 240$ &
Projects the cluster redshift into the same latent space as the GRU representation through a learnable linear transformation. \\
\hline
6 & Concatenation &
$(40 \times 240)+(40 \times 240)$ &
$40 \times 480$ &
Concatenates the GRU latent representation with the redshift embedding for each radial bin, providing both local ICM and global cluster information to the prediction head. \\
\hline
7 & Linear &
$40 \times 480$ &
$40 \times 2$ &
Maps the combined latent representation to two output channels at each radial bin. These correspond to the predicted cumulative mass profile and its associated uncertainty. \\
\hline
\end{tabular}
\label{tab:gru_architecture}
\end{threeparttable}
\end{table*}
\section{GRU model variants with different feature sets}
\setlength{\parskip}{0pt}
\begin{figure}[!htbp]
		\includegraphics[width=0.40\textwidth]{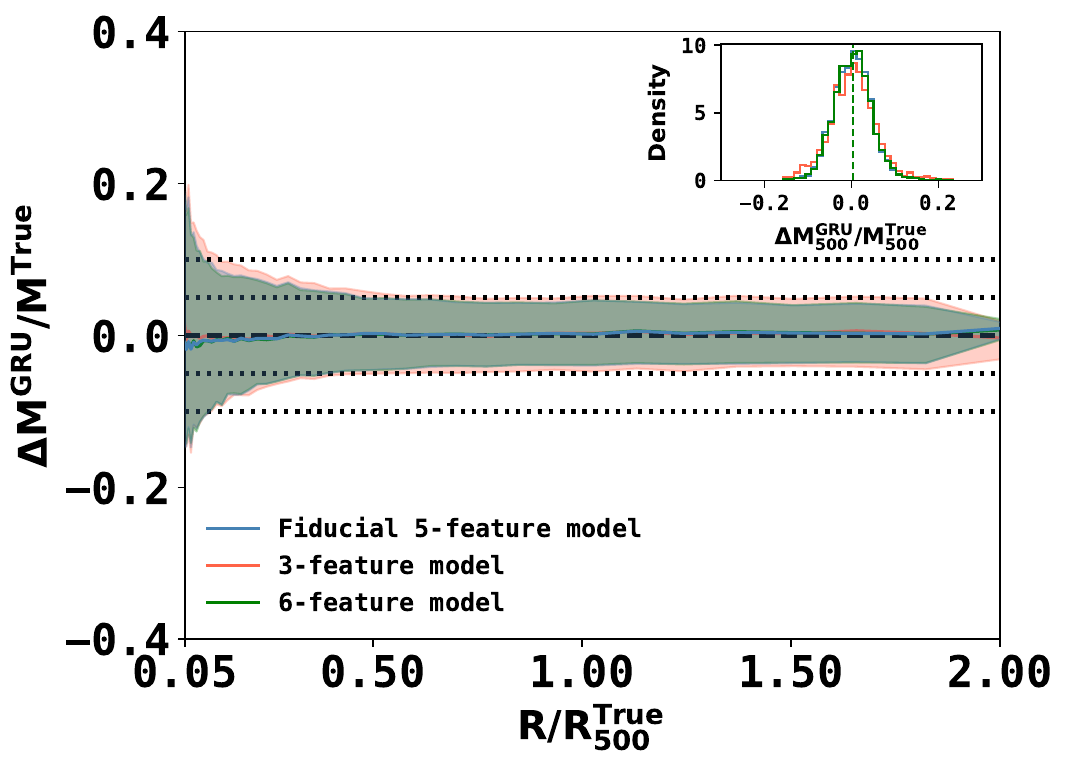}
        \centering
		\caption{\footnotesize Comparison of median fractional residuals of the GRU predictions for the three-feature and six-feature and fiducial five-feature models.  The horizontal dotted black lines represent the  $\pm$5\% and $\pm 10\%$ residuals, while the dashed line represents zero residuals. The inset shows the distribution of fractional residuals in M$_{500}$ for both models, with vertical dashed lines indicating the median of each distribution.
		}
		\label{figcompare236}
\end{figure}
Here, we examine two additional variants of the GRU model, beyond the five-feature and two-feature configurations discussed in the main text, which differ in the number of input features employed. The lower-dimensional three-feature variant uses ICM density and temperature profiles, together with the corresponding radial coordinate. In contrast, the higher-dimensional six-feature variant extends the fiducial five-feature model by including entropy\footnote{The entropy of the ICM is defined as ${\rm K}_{g} = k_b{\rm T}_{g} / n_e^{2/3}$, where k$_{\rm b}$ is the Boltzmann constant and n$_e$ is the electron density.} (K$_g$) as an additional input feature. Figure \ref{figcompare236} presents the median fractional residuals for the two models. The scatter in M$_{500}$ is 0.025 dex for the three-feature model and 0.019 dex for the six-feature model, indicating only marginal differences in predictive accuracy despite the substantial change in input dimensionality. Similarly, SHAP feature importance is  [R, n$_{g}$, T$_{g}$] = [0.25, 0.43, 0.32] and  [R, n$_{g}$, T$_{g}$, P$_{g}$, K$_{g}$, M$_{g}$]=[0.15, 0.09, 0.08, 0.19, 0.09,  0.40]  for the two-feature and six-feature models, respectively. Although entropy exhibits a radial increase similar to that of the gas mass profile, it remains a weak tracer of the underlying mass distribution. 
\section{GRU based mass profile reconstruction from simulated ICM profiles at different resolutions}
\label{mass_reconstruction}
\begin{figure*}[!htbp]
\centering
		\includegraphics[width=.99\textwidth]{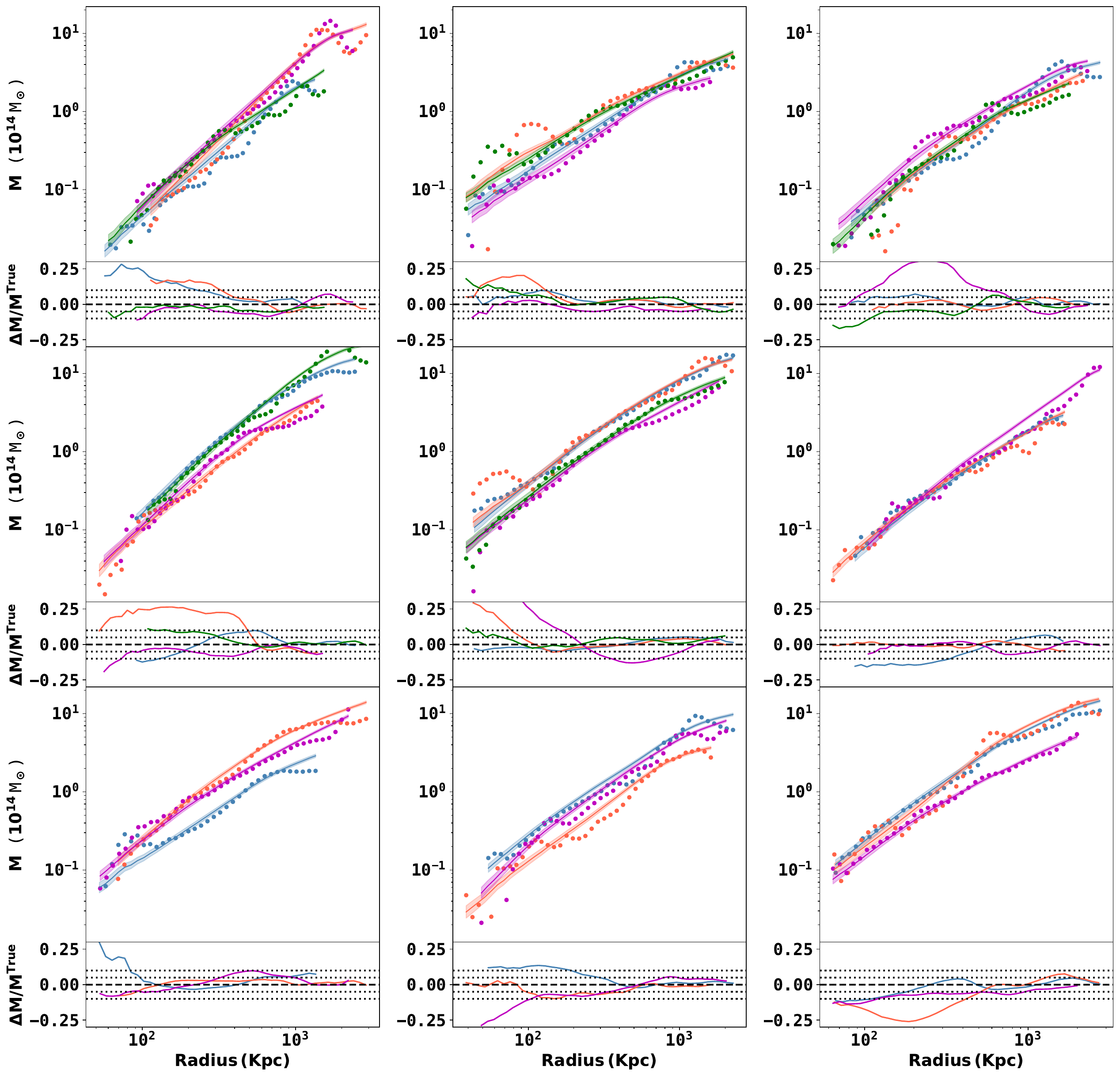}
		\caption{\footnotesize Cumulative mass profile reconstruction of 32 randomly selected clusters using the unaugmented testing sample with the GRU model. The solid lines show the GRU reconstructed profiles, while shaded bands indicate the  16th–84th percentiles. The dotted lines show the mass estimates using the hydrostatic approximation. The lower panels display the fractional residuals (M$^{\rm GRU}$-M$^{\rm True}$)/M$^{\rm True}$. The horizontal dotted black lines represent the  $\pm$5\% and $\pm 10\%$ residuals, while the dashed line represents zero residuals.}
		\label{figA_rec_fine}
\end{figure*}

\begin{figure*}[!htbp]
\centering
		\includegraphics[width=.99\textwidth]{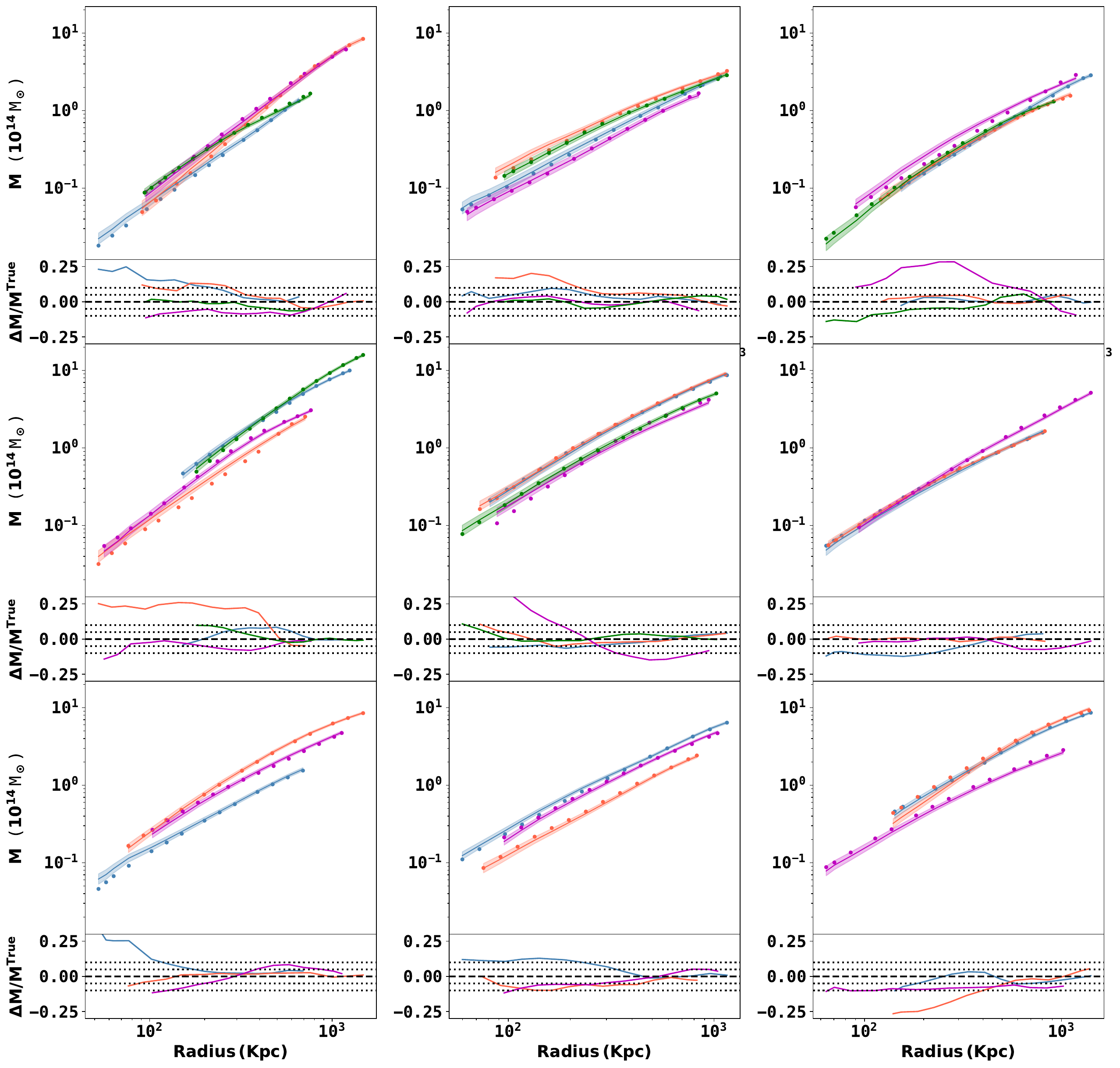}
		\caption{\footnotesize Cumulative mass profile reconstruction of 32 randomly selected clusters using the testing sample with the GRU model using observational-like 14 bins up to 0.75\,R$_{500}$. The solid lines show the GRU reconstructed profiles, while shaded bands indicate the $84^{\rm th}-16^{\rm th}$ percentiles. The dotted points show the true profiles. The lower panels display the fractional residuals (M$^{\rm GRU}$-M$^{\rm True}$)/M$^{\rm True}$. The horizontal dotted black lines represent the  $\pm$5\% and $\pm 10\%$ residuals, while the dashed line represents zero residuals.}
		\label{figA_rec_coarse_1}
\end{figure*}

\begin{figure*}[!htbp]
\centering
		\includegraphics[width=.99\textwidth]{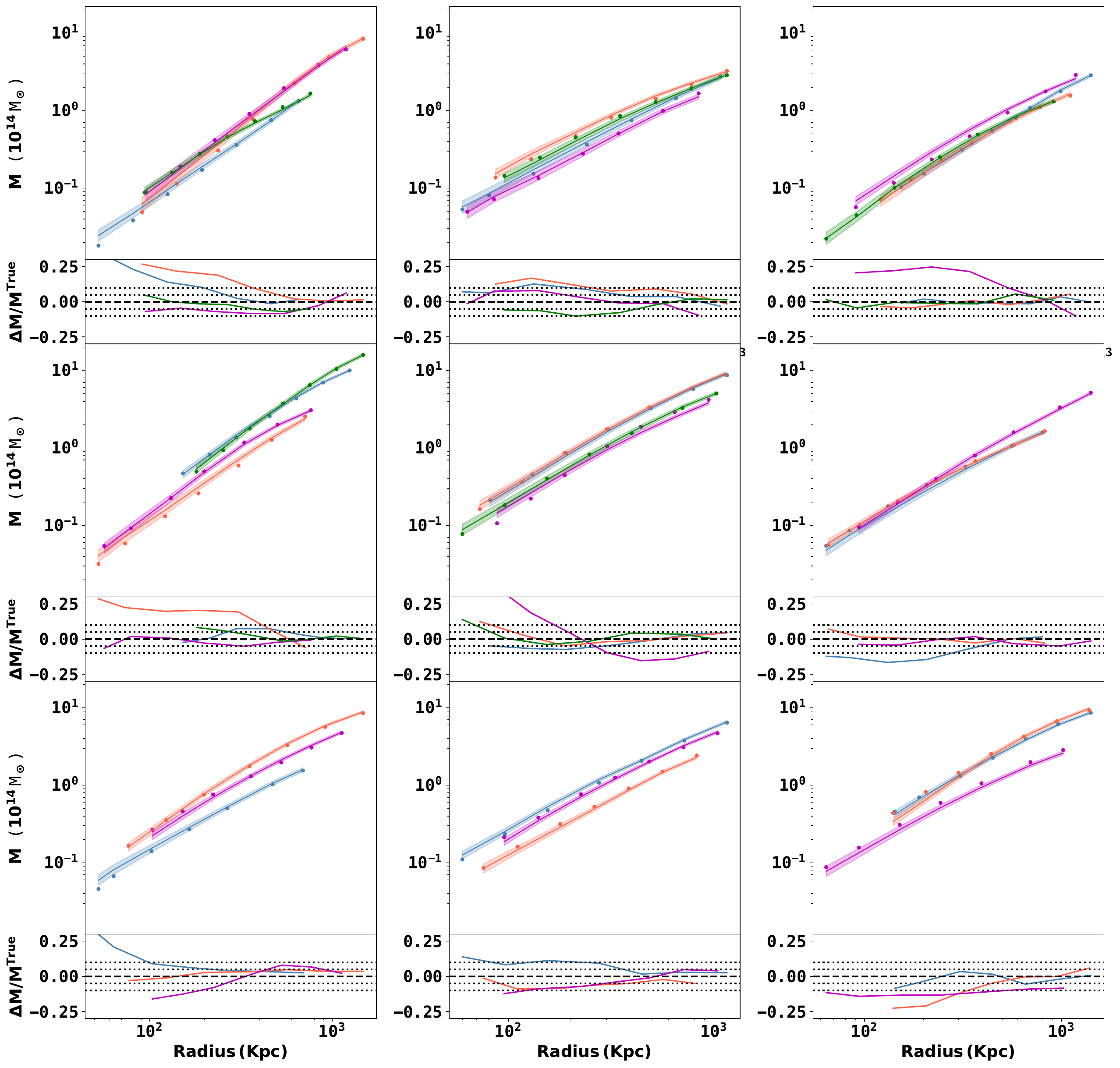}
		\caption{\footnotesize Cumulative mass profile reconstruction of 32 randomly selected clusters using the testing sample with the GRU model using observational-like 7 bins up to 0.75\,R$_{500}$. The solid lines show the GRU reconstructed profiles, while shaded bands indicate the $84^{\rm th}-16^{\rm th}$ percentiles. The dotted points show the true profiles. The lower panels display the fractional residuals (M$^{\rm GRU}$-M$^{\rm True}$)/M$^{\rm True}$. The horizontal dotted black lines represent the  $\pm$5\% and $\pm 10\%$ residuals, while the dashed line represents zero residuals.}
		\label{figA_rec_coarse_2}
\end{figure*}

Figures~\ref{figA_rec_fine}, \ref{figA_rec_coarse_1}, and \ref{figA_rec_coarse_2} present representative GRU mass profile reconstructions for randomly selected 32 clusters from the testing sample using fine radial sampling and observational-like samplings with 14 and 7 radial bins. The examples demonstrate that the GRU model maintains accurate and stable reconstructions across different radial resolutions, while the predicted uncertainty bands become larger as the available radial information is reduced.

\section{Mass profiles reconstruction for the REXCESS and X-COP sample}
\label{appREXCESS}
\begin{figure*}[!htbp]
\centering
		\includegraphics[width=.99\textwidth]{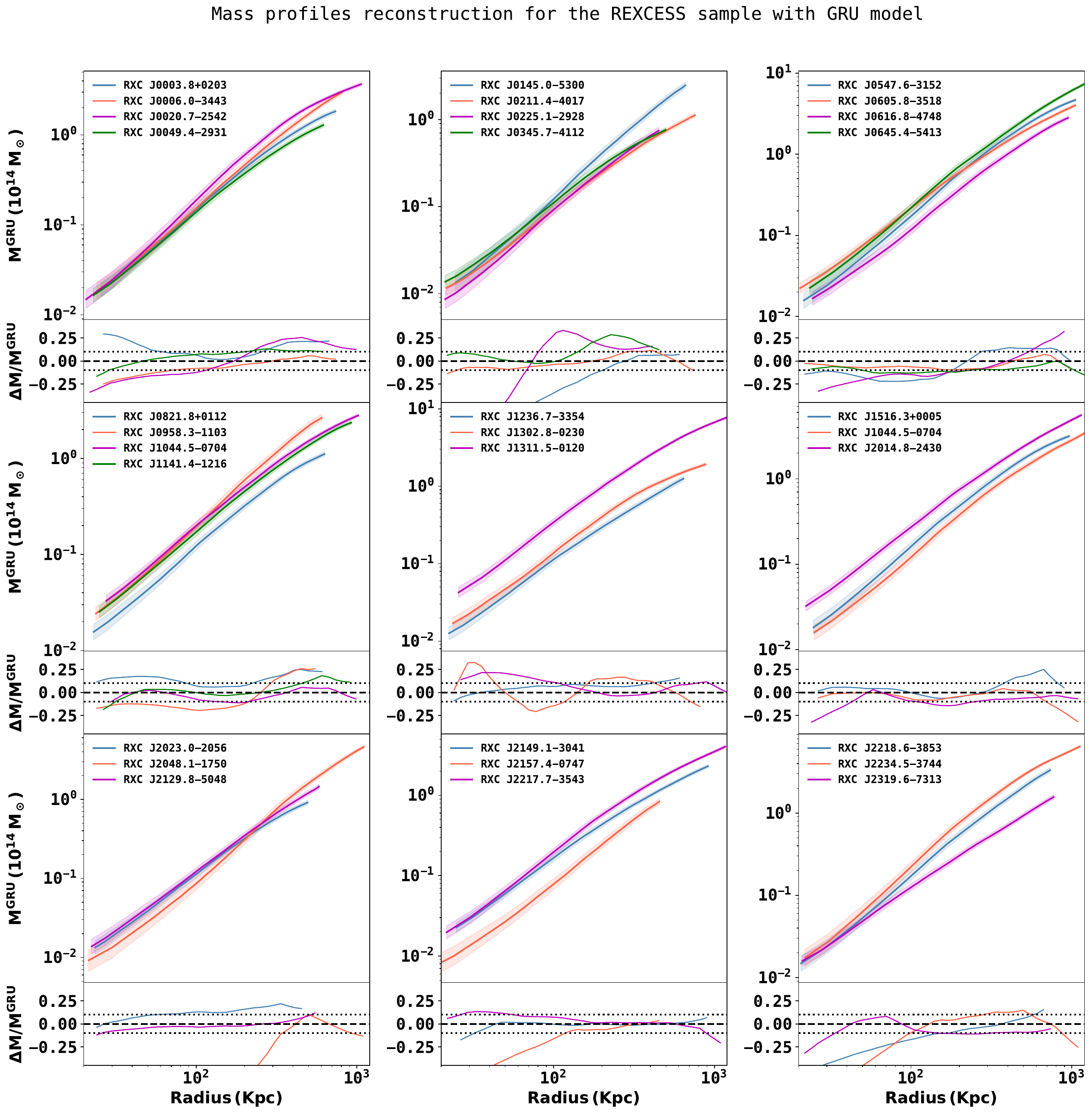}
		\caption{\footnotesize Cumulative mass profile reconstruction of the REXCESS clusters with the GRU model. The shaded bands correspond to the $84^{\rm th}-16^{\rm th}$ of the MC posterior predictive distribution, taking into account both the GRU model uncertainties as well as observational uncertainties in the ICM. The lower panels display the fractional residuals of the hydrostatic mass profiles relative to the GRU based best fit reconstructions, (M$^{\rm HSE}$-M$^{\rm GRU}$)/M$^{\rm GRU}$. Horizontal dotted black lines represent the $\pm 10\%$ residuals, while the dashed line represents zero residuals.}
		\label{figA_REXCESS}
\end{figure*}

\begin{figure*}[!htbp]
\centering
		\includegraphics[width=.99\textwidth]{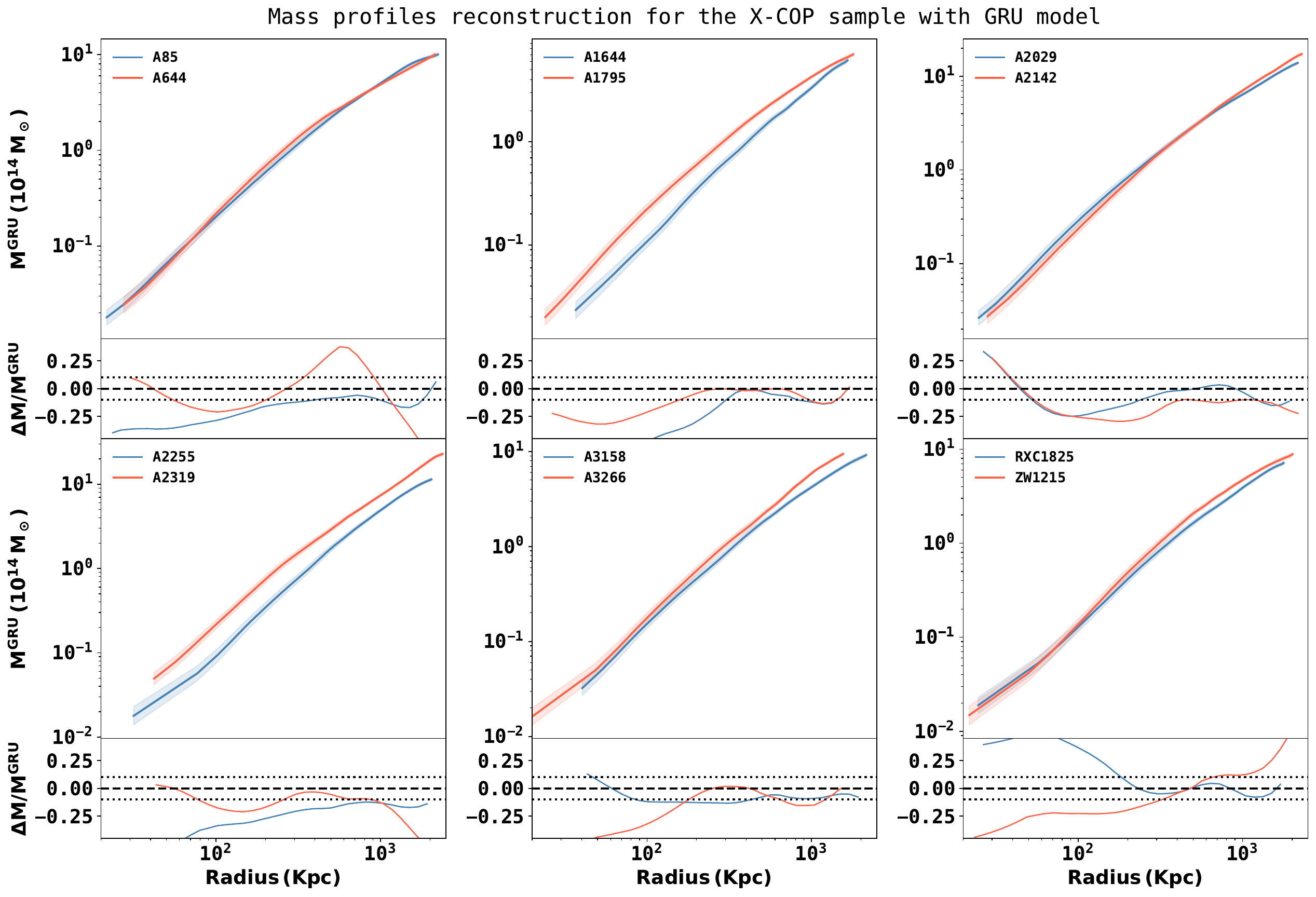}
		\caption{\footnotesize Cumulative mass profile reconstruction of the X-COP clusters with the GRU model. The shaded bands correspond to the $84^{\rm th}-16^{\rm th}$ percentiles of the MC posterior predictive distribution, taking into account both the GRU model uncertainties as well as observational uncertainties in the ICM. The lower panels display the fractional residuals of the hydrostatic mass profiles relative to the GRU based best-fit reconstructions, (M$^{\rm HSE}$-M$^{\rm GRU}$)/M$^{\rm GRU}$. Horizontal dotted black lines represent the $\pm 10\%$ residuals, while the dashed line represents zero residuals.}
		\label{figA_XCOP}
\end{figure*}

\begin{figure}[!htbp]
	\centering
		\includegraphics[width=.4\textwidth]{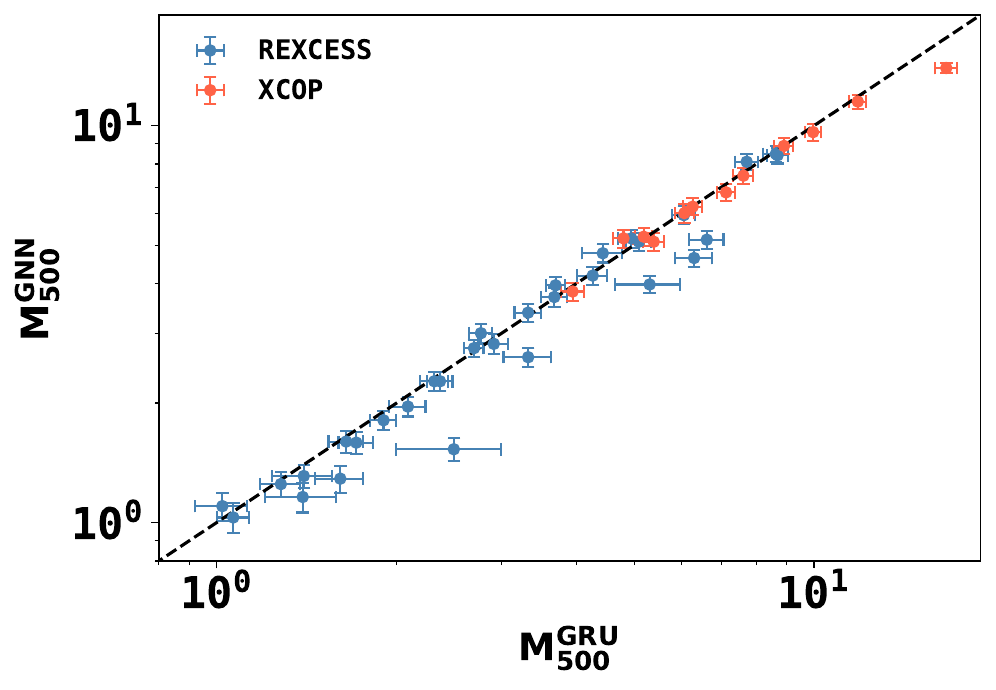}
		\caption{\footnotesize Comparison of M$_{500}$ inferred by the GNN from \cite{Iqbal2025} model (M$^{\rm GNN}_{500}$) and the GRU model (M$^{\rm GRU}_{500}$) developed in this work for the REXCESS (blue) and X-COP (red) galaxy cluster samples. Error bars represent the 1$\sigma$ uncertainties on the predicted masses.}
		\label{figA_XCOPBB}
\end{figure}

Figures~\ref{figA_REXCESS} and \ref{figA_XCOP} show the individual GRU-reconstructed cumulative mass profiles for all clusters in the REXCESS and X-COP samples, respectively, including their associated uncertainty envelopes.  In addition, Fig.~\ref{figA_XCOPBB} shows that the GRU-derived (M$^{\rm GRU}_{500}$) estimates for the REXCESS and X-COP samples are in good agreement with those obtained from the GNN model (M$^{\rm GNN}_{500}$) of \citet{Iqbal2025}. The close agreement between the GNN and GRU predictions provides an additional validation of the GRU model, despite the differences in their training datasets. Tables~\ref{appx1} and \ref{appx2} summarise the cluster mass and sparsity estimates, including the SZ-based masses, GNN-derived M$_{500}$ values, GRU-derived M$_{500}$ and M$_{200}$ estimates, and the corresponding GRU-derived sparsities.

\begin{table}[!htbp]
    \scriptsize
    \centering
\begin{threeparttable}
\setlength{\tabcolsep}{4pt}   
	\caption{\footnotesize Results with REXCESS sample. } 
\begin{tabular}{cccccc}
 \toprule
\toprule
Name & $z$  & M$_{500}^{{\textrm{SZ}}}$\tnote{*} & M$_{500}^{\textrm{GNN}}$\tnote{*}&  M$_{500}^{\textrm{GRU}}$& S$^{\rm GRU}_{500,2500}$ \\
 & & $10^{14}$M$_\odot$ & $10^{14}$M$_\odot$ & $10^{14}$M$_\odot$ &\\
\midrule
RXC J0003.8$+$0203& 0.092 & $2.11^{+0.04}_{-0.04}$    &$2.21^{+0.29}_{-0.29}$ &  $2.08^{+0.14}_{-0.14}$  &   $1.89 ^{+0.19 }_{-0.19 }$ \\[2pt] \hline 
RXC J0006.0$-$3443& 0.114 &$3.82^{+0.29}_{-0.29}$     &$4.77^{+0.25}_{-0.25}$ &  $4.42^{+0.34}_{-0.34}$  &   $2.72 ^{+0.30 }_{-0.30 }$ \\[2pt] \hline 
RXC J0020.7$-$2542& 0.141 &$4.22^{+0.32}_{-0.32}$     &$3.70^{+0.21}_{-0.21}$ &  $3.67^{+0.18}_{-0.18}$  &   $1.75 ^{+0.11 }_{-0.11 }$ \\ [2pt]\hline 
RXC J0049.4$-$2931& 0.108 &$2.14^{+0.36}_{-0.36}$     &$1.60^{+0.10}_{-0.10}$ &  $1.64^{+0.10}_{-0.10}$  &   $1.97 ^{+0.19 }_{-0.19 }$ \\[2pt] \hline 
RXC J0145.0$-$5300& 0.116 &$3.42^{+0.26}_{-0.26}$     &$4.64^{+0.23}_{-0.23}$ &  $6.29^{+0.44}_{-0.44}$  &   $4.07 ^{+0.43 }_{-0.43 }$ \\ [2pt]\hline 
RXC J0211.4$-$4017& 0.100 &$1.48^{+0.32}_{-0.32}$     &$1.03^{+0.09}_{-0.09}$ &  $1.06^{+0.06}_{-0.06}$  &   $2.47 ^{+0.23 }_{-0.23 }$ \\[2pt] \hline 
RXC J0225.1$-$2928& 0.060 &$0.97^{+0.33}_{-0.33}$     &$1.16^{+0.10}_{-0.10}$ &  $1.39^{+0.18}_{-0.18}$  &   $2.69 ^{+0.44 }_{-0.44 }$ \\[2pt] \hline 
RXC J0345.7$-$4112& 0.060 &$1.16^{+0.26}_{-0.26}$     &$1.10^{+0.09}_{-0.09}$ &  $1.02^{+0.10}_{-0.10}$  &   $1.90 ^{+0.21 }_{-0.21 }$ \\[2pt] \hline
RXC J0547.6$-$3152& 0.148 &$4.60^{+0.31}_{-0.31}$     &$5.10^{+0.26}_{-0.26}$ &  $5.09^{+0.24}_{-0.24}$  &   $2.10 ^{+0.13 }_{-0.13 }$ \\[2pt] \hline
RXC J0605.8$-$3518& 0.139 &$5.11^{+0.27}_{-0.27}$     &$4.19^{+0.22}_{-0.22}$ &  $4.26^{+0.24}_{-0.24}$  &   $2.24 ^{+0.17 }_{-0.17 }$ \\[2pt] \hline
RXC J0616.8$-$4748& 0.116 &$2.74^{+0.27}_{-0.27}$     &$2.82^{+0.16}_{-0.16}$ &  $2.91^{+0.16}_{-0.16}$  &   $2.77 ^{+0.26 }_{-0.26 }$ \\[2pt] \hline
RXC J0645.4$-$5413& 0.164 &$7.94^{+0.25}_{-0.25}$     &$8.49^{+0.40}_{-0.40}$ &  $8.63^{+0.41}_{-0.41}$  &   $2.62 ^{+0.21 }_{-0.21 }$ \\[2pt] \hline
RXC J0821.8$+$0112& 0.082 &$2.27^{+0.30}_{-0.30}$     &$1.25^{+0.09}_{-0.09}$ &  $1.28^{+0.09}_{-0.09}$  &   $1.91 ^{+0.19 }_{-0.19 }$ \\[2pt] \hline
RXC J0958.3$-$1103& 0.166 &$5.29^{+0.35}_{-0.35}$     &$3.98^{+0.20}_{-0.20}$ &  $5.30^{+0.66}_{-0.66}$  &   $2.36 ^{+0.33 }_{-0.33 }$ \\[2pt] \hline
RXC J1044.5$-$0704& 0.134 &$3.00^{+0.37}_{-0.37}$     &$2.75^{+0.14}_{-0.14}$ &  $2.69^{+0.10}_{-0.10}$  &   $2.03 ^{+0.12 }_{-0.12 }$ \\[2pt] \hline
RXC J1141.4$-$1216& 0.119 &$2.80^{+0.36}_{-0.36}$     &$2.27^{+0.12}_{-0.12}$ &  $2.36^{+0.11}_{-0.11}$  &   $2.05 ^{+0.15 }_{-0.15 }$ \\[2pt] \hline
RXC J1236.7$-$3354& 0.079 &      -                    &$1.29^{+0.10}_{-0.10}$ &  $1.61^{+0.14}_{-0.14}$  &   $2.73 ^{+0.28 }_{-0.28 }$ \\[2pt] \hline
RXC J1302.8$-$0230& 0.084 &$2.15^{+0.39}_{-0.39}$     &$1.81^{+0.10}_{-0.10}$ &  $1.90^{+0.09}_{-0.09}$  &   $1.90 ^{+0.15 }_{-0.15 }$ \\[2pt] \hline
RXC J1311.5$-$0120& 0.183 &$8.72^{+0.35}_{-0.35}$     &$8.42^{+0.39}_{-0.39}$ &  $8.68^{+0.34}_{-0.34}$  &   $1.91 ^{+0.09 }_{-0.09 }$ \\[2pt] \hline
RXC J1516.3$+$0005& 0.118 &$2.19^{+0.58}_{-0.58}$     &$3.38^{+0.18}_{-0.18}$ &  $3.32^{+0.16}_{-0.16}$  &   $2.05 ^{+0.15 }_{-0.15 }$ \\[2pt] \hline
RXC J1044.5$-$0704& 0.119 &$3.76^{+0.41}_{-0.41}$     &$3.00^{+0.16}_{-0.16}$ &  $2.76^{+0.12}_{-0.12}$  &   $2.55 ^{+0.24 }_{-0.24 }$ \\[2pt] \hline
RXC J2014.8$-$2430& 0.153 &$5.14^{+0.35}_{-0.35}$     &$5.95^{+0.31}_{-0.31}$ &  $6.05^{+0.27}_{-0.27}$  &   $2.27 ^{+0.16 }_{-0.16 }$ \\[2pt] \hline
RXC J2023.0$-$2056& 0.056 &$0.84^{+0.40}_{-0.40}$     &$1.31^{+0.09}_{-0.09}$ &  $1.39^{+0.16}_{-0.16}$  &   $2.01 ^{+0.26 }_{-0.26 }$ \\[2pt] \hline
RXC J2048.1$-$1750& 0.147 &$4.25^{+0.36}_{-0.36}$     &$5.20^{+0.27}_{-0.27}$ &  $4.94^{+0.24}_{-0.24}$  &   $3.73 ^{+0.36 }_{-0.36 }$ \\[2pt] \hline
RXC J2129.8$-$5048& 0.079 &$1.91^{+0.27}_{-0.27}$     &$2.61^{+0.14}_{-0.14}$ &  $3.32^{+0.30}_{-0.30}$  &   $3.64 ^{+0.42 }_{-0.42 }$ \\[2pt] \hline
RXC J2149.1$-$3041& 0.118 &$1.94^{+0.39}_{-0.39}$     &$2.27^{+0.12}_{-0.12}$ &  $2.31^{+0.12}_{-0.12}$  &   $2.29 ^{+0.19 }_{-0.19 }$ \\ [2pt]\hline
RXC J2157.4$-$0747& 0.057 &$1.46^{+0.29}_{-0.29}$     &$1.53^{+0.10}_{-0.10}$ &  $2.49^{+0.49}_{-0.49}$  &   $4.21 ^{+0.93 }_{-0.93 }$ \\[2pt] \hline
RXC J2217.7$-$3543& 0.148 &$4.52^{+0.33}_{-0.33}$     &$3.96^{+0.20}_{-0.20}$ &  $3.69^{+0.13}_{-0.13}$  &   $2.15 ^{+0.12 }_{-0.12 }$ \\ [2pt]\hline
RXC J2218.6$-$3853& 0.141 &$4.01^{+0.34}_{-0.34}$     &$5.16^{+0.27}_{-0.27}$ &  $6.61^{+0.44}_{-0.44}$  &   $2.98 ^{+0.32 }_{-0.32 }$ \\[2pt] \hline
RXC J2234.5$-$3744& 0.151 &$6.87^{+0.27}_{-0.27}$     &$8.11^{+0.37}_{-0.37}$ &  $7.71^{+0.33}_{-0.33}$  &   $2.05 ^{+0.10 }_{-0.10 }$ \\[2pt] \hline
RXC J2319.6$-$7313& 0.098 &$1.78^{+0.29}_{-0.29}$     &$1.59^{+0.10}_{-0.10}$ &  $1.71^{+0.11}_{-0.11}$  &   $3.01 ^{+0.27 }_{-0.27 }$ \\  \midrule
\end{tabular}
\tablefoot{Columns: (1) Cluster name, (2) Redshift, (3) SZ mass, M$_{500}^{\rm SZ}$, derived from the Planck SZ scaling relation, (4) GNN estimated mass, M$_{500}^{\textrm {GNN}}$, (5) GRU-derived mass, M$_{500}^{\textrm {GRU}}$ (this work), (6) GRU-derived sparsity, S$_{500,2500}^{\textrm {GRU}}$ (this work).}
\begin{tablenotes}
\item[*] Derived from \cite{Iqbal2025}.

\end{tablenotes}
\label  {appx1}
\end{threeparttable}
    \end{table}

\begin{table}[!htbp]
    \scriptsize
    \centering
\begin{threeparttable}
\setlength{\tabcolsep}{3.5pt}   
\caption{\footnotesize Results with X-COP sample. }
\begin{tabular}{ccccccccc}
 \toprule
\toprule
Name & $z$  & M$_{500}^{{\textrm{SZ}}}$\tnote{*} & M$_{500}^{\textrm{GNN}}$\tnote{*}&  M$_{500}^{\textrm{GRU}}$& S$^{\rm GRU}_{500,2500}$ &  M$_{200}^{\textrm{GRU}}$& S$^{\rm GRU}_{200,500}$\\
 & & $10^{14}$M$_\odot$ & $10^{14}$M$_\odot$ & $10^{14}$M$_\odot$& & $10^{14}$M$_\odot$ & \\
\midrule
A85     &0.055& $ 4.80^{+0.16}_{-0.16}$&  $6.79^{+0.33}_{-0.33}$ & $  7.12^{+0.24}_{-0.24}$ & $2.49^{+0.09}_{-0.09}$ & $ 9.50^{+0.38 }_{-0.38 }$ &  $1.50^{+0.08}_{-0.08}$  \\[2pt]  \hline
A644    &0.070& $ 5.10^{+0.18}_{-0.18}$ & $6.25^{+0.31}_{-0.31}$ & $  6.27^{+0.22}_{-0.22}$ & $1.26^{+0.11}_{-0.11}$ & $ 9.45^{+0.41 }_{-0.41 }$ &  $1.35^{+0.17}_{-0.17}$  \\[2pt]  \hline 
A1644   &0.047& $ 3.43^{+0.17}_{-0.17}$ & $3.82^{+0.20}_{-0.20}$ & $  3.94^{+0.18}_{-0.18}$ & $3.33^{+0.24}_{-0.24}$ & $ 7.06^{+0.43 }_{-0.43 }$ &  $1.92^{+0.20}_{-0.20}$  \\[2pt] \hline  
A1795   &0.062& $ 4.47^{+0.16}_{-0.16}$ & $5.25^{+0.27}_{-0.27}$ & $  5.19^{+0.17}_{-0.17}$ & $2.09^{+0.07}_{-0.07}$ & $ 7.26^{+0.31 }_{-0.31 }$ &  $1.44^{+0.06}_{-0.06}$  \\[2pt] \hline  
A2029   &0.076& $ 7.20^{+0.20}_{-0.20}$ & $9.63^{+0.46}_{-0.46}$ & $  9.96^{+0.32}_{-0.32}$ & $2.01^{+0.08}_{-0.08}$ & $14.84^{+0.70 }_{-0.70 }$ &  $1.45^{+0.08}_{-0.08}$  \\[2pt] \hline  
A2142   &0.090& $ 8.88^{+0.21}_{-0.21}$ & $11.50^{+0.46}_{-0.46}$& $ 11.84^{+0.38}_{-0.38}$ & $2.65^{+0.09}_{-0.09}$ & $18.57^{+0.83 }_{-0.83 }$ &  $1.52^{+0.07}_{-0.07}$  \\[2pt] \hline  
A2255   &0.080& $ 5.20^{+0.14}_{-0.14}$ & $7.48^{+0.35}_{-0.35}$ & $  7.62^{+0.29}_{-0.29}$ & $5.32^{+0.47}_{-0.47}$ & $12.03^{+0.60 }_{-0.60 }$ &  $1.96^{+0.26}_{-0.26}$  \\[2pt] \hline  
A2319   &0.055& $ 8.72^{+0.13}_{-0.13}$ & $13.97^{+0.39}_{-0.39}$& $ 16.65^{+0.72}_{-0.72}$ & $1.93^{+0.11}_{-0.11}$ & $26.07^{+1.41 }_{-1.41 }$ &  $1.39^{+0.08}_{-0.08}$  \\[2pt] \hline  
A3158   &0.059& $ 4.12^{+0.15}_{-0.15}$ & $5.10^{+0.27}_{-0.27}$ & $  5.39^{+0.22}_{-0.22}$ & $2.79^{+0.13}_{-0.13}$ & $ 8.51^{+0.37 }_{-0.37 }$ &  $1.55^{+0.11}_{-0.11}$  \\[2pt] \hline  
A3266   &0.058& $ 6.56^{+0.12}_{-0.12}$ & $8.89^{+0.42}_{-0.42}$ & $  8.91^{+0.32}_{-0.32}$ & $3.96^{+0.29}_{-0.29}$ & $13.78^{+0.88 }_{-0.88 }$ &  $1.71^{+0.19}_{-0.19}$  \\[2pt] \hline  
RXC1825 &0.065& $ 3.71^{+0.19}_{-0.19}$ & $5.20^{+0.27}_{-0.27}$ & $  4.80^{+0.18}_{-0.18}$ & $2.67^{+0.10}_{-0.10}$ & $ 7.58^{+0.41 }_{-0.41 }$ &  $1.50^{+0.07}_{-0.07}$  \\[2pt] \hline 
wZW1215 &0.076&$4.04^{+0.21}_{-0.21}$ & $6.02^{+0.32}_{-0.32}$   & $  6.07^{+0.22}_{-0.22}$ & $3.04^{+0.22}_{-0.22}$ & $ 8.58^{+0.41 }_{-0.41 }$ &  $1.70^{+0.19}_{-0.19}$  \\   \midrule
\end{tabular}
\tablefoot{Columns: (1) Cluster name, (2) Redshift, (3) SZ mass, M$_{500}^{\rm SZ}$, derived from the Planck SZ scaling relation, (4) GNN-derived mass, M$_{500}^{\textrm {GNN}}$, (5) GRU-derived mass, M$_{500}^{\textrm {GRU}}$ (this work), (6) GRU-derived sparsity, S$_{500,2500}^{\textrm {GRU}}$ (this work), (7) GRU-derived M$_{200}^{\textrm {GRU}}$ (this work), (8) GRU-derived sparsity, S$_{200,500}^{\textrm {GRU}}$ (this work).}
 \label  {appx2}
\begin{tablenotes}
\item[*] Derived from \cite{Iqbal2025}.
\end{tablenotes}

\end{threeparttable}
    \end{table}    

\end{document}